\documentclass[12pt]{article}

\usepackage[utf8]{inputenc}
\usepackage[english]{babel}
\usepackage{graphicx}
\usepackage{float}
\usepackage{mathtools}
\usepackage{amsfonts}
\usepackage{amssymb}
\usepackage{bbm}
\usepackage{bm}
\usepackage{tikz}
\usepackage{algpseudocode}
\usepackage{breqn}
\usepackage[hyphens]{url}
\usepackage{etoolbox}
\usepackage{lmodern}
\usepackage[round]{natbib}
\usepackage{tikz}
\usepackage{hyperref}
\mathtoolsset{showonlyrefs}

\usepackage{color-edits}

\usepackage{tikz}
\usetikzlibrary{positioning,shapes.geometric}

\usepackage[parfill]{parskip}

\newtheorem{proposition}{Proposition}
\newtheorem{lemma}{Lemma}
\newtheorem{theorem}{Theorem}

\newtheorem{corollary}{Corollary}
\newtheorem{assumption}{Assumption}

\newtheorem{remark}{Remark}

\newtheorem{definition}{Definition}
\newtheorem{example}{Example}

\usetikzlibrary{shapes,decorations,arrows,calc,arrows.meta,fit,positioning}
\tikzset{
    -Latex,auto,node distance =1 cm and 1 cm,semithick,
    state/.style ={ellipse, draw, minimum width = 0.7 cm},
    point/.style = {circle, draw, inner sep=0.04cm,fill,node contents={}},
    bidirected/.style={Latex-Latex,dashed},
    el/.style = {inner sep=2pt, align=left, sloped}
}

\DeclareRobustCommand{\Rb}{\mathbb{R}}

\DeclareRobustCommand{\bZ}{\boldsymbol{Z}}
\DeclareRobustCommand{\bA}{\boldsymbol{A}}
\DeclareRobustCommand{\bH}{\boldsymbol{H}}
\DeclareRobustCommand{\bE}{\boldsymbol{E}}
\DeclareRobustCommand{\bU}{\boldsymbol{U}}
\DeclareRobustCommand{\bV}{\boldsymbol{V}}

\DeclareRobustCommand{\bX}{\boldsymbol{X}}

\newcommand{\bY}{\boldsymbol{Y}}

\DeclareRobustCommand{\Ac}{\mathcal{A}}

\DeclareRobustCommand{\Hc}{\mathcal{H}}

\DeclareRobustCommand{\Ic}{\mathcal{I}}
\DeclareRobustCommand{\Ec}{\mathcal{E}}

\DeclareRobustCommand{\Rc}{\mathcal{R}}
\DeclareRobustCommand{\Mc}{\mathcal{M}}

\newcommand{\Xc}{\mathcal{X}}

\newcommand{\Tc}{\mathcal{T}}

\DeclareRobustCommand{\Ac}{\mathcal{A}}

\DeclareRobustCommand{\Nc}{\mathcal{N}}

\DeclareRobustCommand{\Rb}{\mathbb R}
\DeclareRobustCommand{\Nb}{\mathbb N}

\DeclareRobustCommand{\norm}[1]{\left\lVert #1 \right\rVert}

\DeclareMathOperator{\1}{\mathbbm 1}

\newcommand{\distas}[1]{\mathbin{\overset{#1}{\kern\z@\sim}}}%
\newsavebox{\mybox}\newsavebox{\mysim}

\newcommand{\distras}[1]{%
  \savebox{\mybox}{\hbox{\kern3pt$\scriptstyle#1$\kern3pt}}%
  \savebox{\mysim}{\hbox{$\sim$}}%
  \mathbin{\overset{#1}{\kern\z@\resizebox{\wd\mybox}{\ht\mysim}{$\sim$}}}%
}

\newcommand{\hbeta}{\widehat{\beta}}

\newcommand{\Ex}{\mathbb{E}}
\newcommand{\Pb}{\mathbb{P}}

\newcommand{\bbeta}{\boldsymbol{\beta}}

\newcommand{\ldot}[2]{\left\langle #1, #2 \right\rangle}

\makeatletter
\newcommand*{\indep}{%
  \mathbin{%
    \mathpalette{\@indep}{}%
  }%
}
\newcommand*{\nindep}{%
  \mathbin{
    \mathpalette{\@indep}{/}%
  }%
}
\newcommand*{\@indep}[2]{%
  \sbox0{$#1\perp\m@th$}
  \sbox2{$#1=$}
  \sbox4{$#1\vcenter{}$}
  \rlap{\copy0}
  \dimen@=\dimexpr\ht2-\ht4-.2pt\relax
  \kern\dimen@
  \ifx\\#2\\%
  \else
    \hbox to \wd2{\hss$#1#2\m@th$\hss}%
    \kern-\wd2 %
  \fi
  \kern\dimen@
  \copy0 
}
\makeatother

\title{
A Causal Inference Framework for Data Rich Environments
}

\author{Alberto Abadie, Anish Agarwal, Devavrat Shah}

\date{December 28, 2023}

\allowdisplaybreaks

\makeatletter
\def\monthname{\ifcase\month\or
January\or February\or March\or April\or May\or June\or
July\or August\or September\or October\or November\or December\fi}
\makeatother

\makeatletter
\def\@sect#1#2#3#4#5#6[#7]#8{\ifnum #2>\c@secnumdepth
     \let\@svsec\@empty\else
     \refstepcounter{#1}\edef\@svsec{\csname the#1\endcsname. \hskip 0.4em}\fi
     \@tempskipa #5\relax
      \ifdim \@tempskipa>\z@
        \begingroup #6\relax
          \@hangfrom{\hskip #3\relax\@svsec}{\interlinepenalty \@M #8\par}%
        \endgroup
      \csname #1mark\endcsname{#7}\addcontentsline
         {toc}{#1}{\ifnum #2>\c@secnumdepth \else
                      \protect\numberline{\csname the#1\endcsname}\fi
                    #7}\else
        \def\@svsechd{#6\hskip #3\relax  
                  \@svsec #8\csname #1mark\endcsname
                      {#7}\addcontentsline
                          {toc}{#1}{\ifnum #2>\c@secnumdepth \else
                             \protect\numberline{\csname the#1\endcsname}\fi
                      #7}}\fi
     \@xsect{#5}}
\makeatother

\makeatletter
\renewcommand{\section}{\@startsection{section}{1}{0mm}{-\baselineskip}{0.25\baselineskip}{\center\normalfont\normalsize\bf}}
\renewcommand{\subsection}{\@startsection{subsection}{2}{0mm}{-\baselineskip}{0.05\baselineskip}{\raggedright\normalfont\normalsize\bf}}
\renewcommand{\subsubsection}{\@startsection{subsubsection}{3}{0mm}{-\baselineskip}{0.05\baselineskip}{\raggedright\normalfont\normalsize\itshape}}
\def\@begintheorem#1#2{\trivlist \item[\hskip \labelsep{\bf #1\ #2:}]\it}
\makeatother

\begin{document}

\maketitle

\begin{abstract}
We propose a formal model for counterfactual estimation with unobserved confounding in ``data-rich'' settings, i.e., where there are a large number of units and a large number of measurements per unit.
Our model provides a bridge between the structural causal model view of causal inference common in the graphical models literature with that of the latent factor model view common in the potential outcomes literature.
We show how classic models for potential outcomes and treatment assignments fit within our framework.
We provide an identification argument for the average treatment effect, the average treatment effect on the treated, and the average treatment effect on the untreated. 
For any estimator that has a fast enough estimation error rate for a certain nuisance parameter, we establish it is consistent for these various causal parameters. 
We then show principal component regression is one such estimator that leads to consistent estimation, and we analyze the minimal smoothness required of the potential outcomes function for consistency.

\end{abstract}


\setcounter{page}{1}
\addtolength{\baselineskip}{0.5\baselineskip}

\section{Introduction}
One of the central goals of empirical economic research is to ascertain the effects of treatments (policies, treatments) on the outcomes of interest. 
A fundamental challenge for the estimation of treatment effects is the pervasive presence of unobserved confounders. 
For example, in a study of the effects of health insurance on healthcare utilization, unobserved or latent health determinants may differ between insured and uninsured individuals, biasing treatment effect estimates. 
Several approaches have been put forward to estimate treatment effects in the presence of confounders, including explicit randomization of the treatment, controlling for measured confounders, and instrumental variable methods. 
Traditionally, these methods are not designed to operate in data-rich environments where the curse of dimensionality creates challenges for estimation and inference, and do not  take advantage of the information contained in high-dimensional data to identify treatment effects.

In recent times, the availability of high-dimensional data on economic behavior has become commonplace. 
Modern data harvesting technologies, based on digitization and pervasive sensors, enable the collection of detailed high-frequency attribute and outcome information on individuals (or other observational units; e.g., geo-locations) concurrently undergoing different treatments. 
For example, electronic health records contain rich information about patients' medical history over time. Similarly, internet retailers and marketing firms use scanner data to collect high-dimensional information on customers' purchases. 
The goal of this article is to provide a framework for causal inference that takes advantage of modern data-rich environments to counter the effect of unobserved or latent confounders. 

Given this goal, we consider a setting where we have access to data for $N$ units (e.g., individuals, sub-populations, firms, geographic locations) and $T$ measurements of outcomes per unit. 
Different measurements may represent the same outcome metric at different time periods, different outcome metrics (e.g., customers' expenditures in different product categories) for the same time period, or a combination of both. 
We argue that (high-dimensional) data-rich environments---i.e., large $N$ and large $T$---make it possible to estimate treatment effects in the presence of unobserved or latent confounding, without needing to make parametric assumptions in the manner in which unobserved confounders affect selection for treatment and the outcome metrics.

\subsection{Contributions and Related Work}
{\bf Contributions.}
We propose a formal model for counterfactual estimation with unobserved confounding in ``data-rich'' settings, i.e., where there are a large number of units and a large number of measurements per unit. 
We posit a general data-generating process (DGP) for how potential outcomes are defined and how treatments are assigned, allowing for unobserved confounding. 
We provide a structural causal model view of the conditional independence conditions required in our DGP that imply that the treatment assignments are exogenous of the potential outcomes conditional on these unobserved confounders. 
We establish that if the unobserved confounders are low-dimensional, relative to the number of units and measurements, and the potential outcomes are a smooth non-linear function of them, it {\em implies} that an approximate linear latent factor model of appropriate dimension holds, where the approximation error decays as the number of units and measurements increase. 
In doing so, we believe this model provides a formal bridge between the structural causal model view of causal inference common in the graphical models literature with that of the latent factor model view common in the potential outcomes literature.

We formalize how classic models for potential outcomes and treatment assignments fit within our framework. 
For the potential outcomes, we show how two-way fixed effects, interactive fixed effects, binary choice, and dictionary basis expansions fit within our framework. 
For treatment assignments, we show how randomized control trials (RCTs), selection on (un)observables, regression discontinuity, random utility models, and staggered adoption settings fit within our framework.
Theoretically, we provide an identification argument for the average treatment effect (ATE), the average treatment effect on the treated (ATT), and the average treatment effect on the untreated (ATU). 
For any estimator that has a fast enough estimation error rate for a certain nuisance parameter, we establish it is consistent for these various causal parameters. 
We then show principal component regression (PCR) is one such estimator that leads to consistent estimation, and we analyze the minimal smoothness required of the potential outcomes function for consistency.

{\bf Related work.}
This model builds upon the latent factor model literature studied in the growing literature on causal panel data models (\citep{chamberlain_factor, SC, bai2009interactive, athey2021matrix, bai2021matrix, SDID, CMC, SI, dwivedi2022doubly, synth_combo}).
The model we propose can be viewed as a generalization of the exact linear factor model studied in these works to a non-linear factor model. 
Our model allows for both panel data and cross-sectional data, and combinations thereof.
Importantly, we argue that beginning from a general structural causal model, if the outcomes are a smooth function of the unobserved confounders, then it {\em implies} that a factor model of appropriate dimension holds if there are large number of units and measurements, thereby hopefully providing a bridge between these two frameworks.
In terms of the estimator we propose, to the best of our knowledge, this is also the first theoretical analysis of consistency for the ATE with a (smooth) non-linear factor model. 
It is also the first analysis of PCR (\cite{PCR1, PCR2, CICD, PCR3}) for such target causal estimands with unobserved confounding.
This requires dealing with the novel technical challenge of error-in-variables, only an approximate low-rank model holding on the noiseless covariates, and linear misspecification error.

\subsection{Notation}
For a matrix $\bA \in \Rb^{a \times b}$, we denote its transpose as $\bA^T \in \Rb^{b \times a}$. 
We denote the operator (spectral) and Frobenius norms of $\bA$ as $\|\bA\|_{\text{op}}$ and $\|\bA\|_F$, respectively. 
The columnspace (or range) of $\bA$ is the span of its columns, which we denote as  $\Rc(\bA) = \{v \in \Rb^a: v = \bA x, x \in \Rb^b\}$. 
The rowspace of $\bA$, given by $\Rc(\bA^T)$, is the span of its rows. 
Recall that the nullspace of $\bA$ is the set of vectors that are mapped to zero under $\bA$. 
For any vector $v \in \Rb^a$, let $\|v\|_p$ denote its $\ell_p$-norm, and let $\|v\|_\infty$ denote its max-norm. 
The inner product between vectors $v, x \in \Rb^a$ is $\langle v, x \rangle = \sum_{\ell=1}^a v_\ell x_\ell$. 
If $v$ is a random variable, we denote its sub-Gaussian (Orlicz) norm as $\|v\|_{\psi_2}$. For any positive integer $a$, we use the notation $[a] = \{1, \dots, a\}$. 

Let $f$ and $g$ be two real-valued functions defined on $\mathcal X$, an unbounded subset of $[0,\infty)$. 
We say that $f(x)$ = $O(g(x))$ if and only if there exists a positive real number $M$ and $x_0\in \mathcal X$ such that, for all $x \ge x_0$, we have $|f (x)| \le M|g(x)|$. 
Analogously, we say that $f (x) = \Theta(g(x))$ if and only if there exist positive real numbers $m, M$ and $x_0\in \mathcal X$ such that for all $x \ge x_0$, we have $m|g(x)| \le |f(x)| \le M|g(x)|$; $f (x) = o(g(x))$ if for any $m > 0$, there exists $x_0\in \mathcal X$ such that for all $x \ge x_0$, we have $|f(x)| \le m|g(x)|$.

We adopt standard notation and definitions for stochastic convergence. 
We employ $\xrightarrow{d}$ and $\xrightarrow{p}$ to indicate convergence in distribution and probability, respectively. 
For any sequence of random vectors, $X_n$, and any sequence of positive real numbers, $a_n$, we say $X_n = O_p(a_n)$ if for every $\varepsilon>0$, there exists constants $C_\varepsilon$ and $n_\varepsilon$ such that $\Pb( \| X_n \|_2 > C_\varepsilon a_n) < \varepsilon$ for every $n \ge n_\varepsilon$; equivalently, we say $(1/a_n) X_n$ is uniformly tight or bounded in probability. 
$X_n = o_p(a_n)$ means $X_n/a_n\xrightarrow{p} 0$. We say a sequence of events $\Ec_n$, indexed by $n$, holds ``with probability approaching one'' (w.p.a.1) if $\Pb(\Ec_n) \rightarrow 1$ as $n \rightarrow \infty$, i.e., for any $\varepsilon > 0$, there exists a $n_\varepsilon$ such that for all $n > n_\varepsilon$, $\Pb(\Ec_n) > 1 - \varepsilon$. 
More generally, a multi-indexed sequence of events $\Ec_{n_1,\dots, n_d}$, with indices $n_1,\dots, n_d$ with $d \geq 1$, is said to hold w.p.a.1 if $\Pb(\Ec_{n_1,\dots, n_d}) \rightarrow 1$ as $\min\{n_1,\dots, n_d\} \rightarrow \infty$. 
We also use $\Nc(\mu, \sigma^2)$ to denote a normal or Gaussian distribution with mean $\mu$ and variance $\sigma^2$---we call it {\em standard} normal or Gaussian if $\mu = 0$ and $\sigma^2 = 1$.
We use $C$ to denote a positive constant, with a value that can change across instances. 

\section{Model}\label{sec:model}
We are interested in evaluating the effect of treatments on outcomes of interest. 
Specifically, we observe $T$ outcomes or measurements for $N$ units. 
For each measurement $t \in [T]$ and unit $n \in [N]$, we observe $Y_{n, t} \in \Rb$ under treatment $A_{n, t} \in \Ac$, where $|\Ac| = A$.

Let $\bA = [A_{n, t}]_{n \in [N], t \in [T]} \in \Ac^{N \times T}$ and $\bY = [Y_{n, t}]_{n \in [N], t \in [T]} \in \Rb^{N \times T}$ collect the matrix of treatment assignments and outcomes, respectively.
We now define how the treatment assignments and outcomes are generated.
We define the random variables $\bU \in \mathcal{U}, \bE \in \mathcal{E}$ and functions  $h: \mathcal{U} \to \Ac^{N \times T}$, $f: \Ac^{N \times T} \times \mathcal{U} \times \mathcal{E}  \to \Rb^{N \times T}$ such that
\begin{align}
\bA &= h(\bU) \label{eq:basic_DGP_1},
\\ \bY &= f(\bA, \bU, \bE) \label{eq:basic_DGP_2}.
\end{align}
We allow the functions $h, f$ and the variables $\bU, \bE$ to be unobserved; that is, we only observe the treatment assignments $\bA$ and the outcomes $\bY$.
$\bU$ contains all potential confounders that can affect both the treatment assignment $\bA$ and the outcomes $\bY$.
$\bE$ is the random variation in $\bY$ not explained by $\bU$.
The question we study in this paper is:

\begin{center}
{\em As $N$ and $T$ grow, under what conditions on the outcomes---the function $f$ and the unobserved variables $\bU, \bE$---is effective counterfactual inference possible despite unobserved confounding?}
\end{center}

\subsection{Data Generating Process}
Towards answering the question above, we assume the following data-generating process (DGP).
As stated earlier, we hope this DGP serves a bridge between the SCM and latent factor model view of causal inference. 

\begin{assumption}[Data generating process]\label{assumption:DGP}{\color{white}.}
\begin{enumerate}
\item We assume the following factorization of $\bU, \bE$:
\begin{align}
\bU = [U_n]_{n \in [N]}, \quad \bE = [\varepsilon^{(a)}_{n, t}]_{n \in [N], t \in [T], a \in \Ac}
\end{align}
where $U_n \in \Rb^q$, and $\varepsilon^{(a)}_{n, t} \in \Rb$ for some $q \geq 1$.

\item We do not make any distributional assumptions about $\bU$ and it can be thought to be conditioned on for the remainder of the paper.
Conditional on $\bU$, for all $n\in [N]$ and $t\in [T]$, we assume the vector $(\varepsilon^{(a)}_{n, t})_{a \in \Ac}$ is sampled independently.
Hence, the only source of uncertainty in our model is due to $\bE$.

\item We assume $f$ has the following factorization: for $n \in [N], t \in [T]$, potential and observed outcomes are generated as 
\begin{align}\label{eq:potential_outcomes_function}
Y^{(a)}_{n, t} &= f_{t, a}\Big(U_n\Big) + \varepsilon^{(a)}_{n, t}, \text{ for } a \in \Ac,
\\ Y_{n, t} &= Y^{(A_{n,t})}_{n, t},
\end{align}
where $f_{t, a}: \Rb^q \to \Rb$, $\varepsilon^{(a)}_{n, t} \in \Rb$, and we assume $\Ex[\varepsilon^{(a)}_{n, t} \mid \bU] = 0$. 
$\varepsilon^{(a)}_{n, t}$ can be interpreted as capturing the random variation in the potential outcomes $Y^{(a)}_{n, t}$ that is not captured by $f_{t, a}$.  
As discussed in Section \ref{sec:latent_strucutre_outcomes}, various models for potential outcomes considered in the literature can be captured via Eq \eqref{eq:potential_outcomes_function} as long as $f_{t, a}$ is sufficiently smooth.
\end{enumerate}

\end{assumption}

\begin{figure}[h!]
\begin{center}
\begin{tikzpicture}
\node (A) at (0,0) {$A_{n,t}$};
\node (U) at (1,1) {$U_n$};
\node (Y) at (2,0) {$Y_{n,t}$};
\node (varepsilon) at (3,1) {$(\varepsilon^{(a)}_{n, t})_{a \in \Ac}$};
\draw[thick, ->] (U) -- (A);
\draw[thick, ->] (U) -- (varepsilon);
\draw[thick, ->] (U) -- (Y);
\draw[thick, ->] (A) -- (Y);
\draw[thick, ->] (varepsilon) -- (Y);
\end{tikzpicture}\vspace*{-.5cm}
\end{center}
\caption{DAG representation of data generating process.}
\label{figure:dag}
\end{figure}
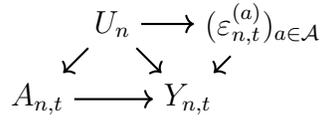

\begin{remark}
The DAG in Figure \ref{figure:dag} is consistent with the independence assumptions we make above in the DGP. 
Further, Assumption \ref{assumption:DGP} implies the following conditional exogeneity condition
\begin{align}
Y^{(a)}_{n, t} \indep \bA \mid U_n.
\end{align}
\end{remark}

\begin{remark}\label{remark:non-linear_epsilon}
A potential outcome function of the form $Y^{(a)}_{n, t} = \bar{f}_{t, a}(U_n, \bar{\varepsilon}^{(a)}_{n, t})$ can be nested into \eqref{eq:potential_outcomes_function} under additional conditions on the distribution of $\bar{\varepsilon}^{(a)}_{n, t}$.
In particular, assume that the distribution of $\bar{\varepsilon}^{(a)}_{n, t}$ is independent of $n$, conditional on $\bU$. 
Then $$\Ex[Y^{(a)}_{n, t} \mid \bU] = \Ex[\bar{f}_{t, a}(U_n, \bar{\varepsilon}^{(a)}_{n, t}) \mid \bU] = f_{t,a}(U_n),$$ where the expectation is taken with respect to $\bar{\varepsilon}^{(a)}_{n, t}$.
In particular, because the distribution of $\bar{\varepsilon}^{(a)}_{n, t}$ is not dependent on $n$, the conditional expectation $\Ex[Y^{(a)}_{n, t} \mid U_n]$ can be written as only a function of $U_n$, and $t, a$.
Then by defining $\varepsilon^{(a)}_{n, t} = Y^{(a)}_{n, t} - \Ex[Y^{(a)}_{n, t} \mid U_n],$
\eqref{eq:potential_outcomes_function} holds.
\end{remark}

\section{Outcome and Treatment Assignment Functions Within our Framework}\label{sec:latent_structure}
Thus far, the setup has been quite general. 
To make progress, we impose relatively generic smoothness conditions on $f_{t, a}$, and argue this encompasses familiar models for potential outcomes considered in the econometric literature.
Further, we show how various models for the treatment assignment functions studied in the literature can be encompassed within our framework.

\subsection{Outcome Functions}\label{sec:latent_strucutre_outcomes}

We first formally define what we mean by smoothness.

\begin{definition}[H\"older continuity, e.g., \citeauthor{jiaming_spectral_convergence}, 
\citeyear{jiaming_spectral_convergence}]\label{def:holder}
For $k \geq 1$, let $s = (s_1, \dots, s_k)$ be a $k$-tuple of non-negative integers with $|s| = \sum^k_{\ell = 1} s_\ell$. For $S \in \Nb$ and $C_H>0$, the H\"older class $\Hc(k,S,C_H)$ on $[0, 1)^k$ is the set of functions $g: [0, 1)^k \to \mathbb{R}$ with partial derivatives that satisfy
\[
\sum_{s: |s| = S - 1} \frac{1}{s!} |\nabla_{\!s} g(\mu) - \nabla_{\!s} g(\mu') | \le C_H \norm{\mu - \mu'}_{\infty},\quad \forall \mu, \mu' \in [0, 1)^k.
\footnote{Note that for any compact set $\Xc \in \Rb^k$, we have $\Xc \subset [-c, c)^k$, where $c \le \infty$. 
Then, $[-c, c)^k$ can be replaced by $[0, 1)^k$, without loss of generality by re-scaling.}
\]
\end{definition}
In essence, Definition \ref{def:holder} requires that the $(S-1)$-th derivatives of $g$ are Lipchitz continuous.
For example, it is easy to verify that an analytic function with compact domain is H\"older continuous for all $S \in \Nb$.

\begin{assumption}\label{assumption:holder.mech}
For all $n \in [N]$, recall $U_n \in [0,1)^q$.
For all $t \in [T], a \in \Ac$, we assume $f_{t, a}$ is H\"older continuous, i.e., $f_{t, a} \in \Hc(q,S,C_H)$, where $C_H < C < \infty$. 
\end{assumption}
Informally, Assumption \ref{assumption:holder.mech} is a continuity condition that posits that if latent variables $U_{n_1}$ and $U_{n_2}$ for any two units $n_1$ and $n_2$ are close ($U_{n_1}\approx U_{n_2}$), then their average potential outcomes are close as well, ($\Ex[Y^{(a)}_{n_1, t}] \approx \Ex[Y^{(a)}_{n_2, t}]$, for all $t \in [T], a \in \Ac$), where the expectation is taken with respect to $\varepsilon^{(a)}_{n_1, t}$ and $\varepsilon^{(a)}_{n_2, t}$, respectively. 

A linear factor model, $f_{t, a}(U_n) = \ldot{U_n}{\tilde{U}_{t, a}}$, is a special case of Assumption \ref{assumption:holder.mech} and one can verify it satisfies Definition \ref{def:holder} for all $S \in \mathbb{N}$ .
Proposition \ref{prop:factor_model_approx} establishes that linear factor models of sufficiently large dimension also provide a ``universal'' representation for smooth non-linear factor models.

\begin{proposition}[H\"older low rank matrix approximation,   \citeauthor{jiaming_spectral_convergence}, 
\citeyear{jiaming_spectral_convergence}]\label{prop:factor_model_approx}
Suppose Assumption \ref{assumption:holder.mech} holds.
Then, for all $n \in [N], t \in [T], a \in \Ac$, and any $\delta>0$, there exist latent variables $\lambda_n, \rho_{t, a} \in \Rb^r$ such that:
\begin{align}
\left| f_{t, a}(U_n) - \ldot{\lambda_n}{\rho_{t, a}} \right| \leq \Delta_E,
\end{align}
where for $\bar{C}$ that is allowed to depend on $(q,S)$,
\begin{align}
\Delta_E \le C_H \cdot \delta^S\quad\mbox{with}\quad r \le \bar{C} \cdot \delta^{-q}.
\end{align}
\end{proposition}
Proposition \ref{prop:factor_model_approx} establishes that if $f_{t, a}$ has a H\"older smooth latent variable representation, then it is uniformly well-approximated by a linear factor model of finite dimension, $r$.
For $\delta < 1$ we have that as the latent dimension $q$ of the confounder $U_n$ increases, the bound on the rank $r$ increases, and as smoothness $S$ of $f_{t, a}$ increases, the bounds on the approximation error $\Delta_E$ decreases.
If we take $\delta = (\mbox{min}\{N, T\})^{-c}$ for some constant $c$, such that $0<c<1/q$, we obtain $r \ll \mbox{min}\{N, T\}$ and $\Delta_E = o(1)$ as $N, T \to \infty$.

\subsubsection{Classic Econometric Models that Fit our Framework}\label{ssec:examples}

Our framework nests some classical econometric models, which we describe below.

\begin{example}[Two-way fixed effects model]\label{example:two_way_FE}
Suppose 
\[
Y^{(a)}_{n, t} = \ldot{a}{\beta_n} + \mu_n + w_t + \varepsilon^{(a)}_{n, t}
\]
Here, $\mu_n$, $v_t$ are scalars, and $a$ and $\beta_n$ have dimension $p$. This corresponds to a two-way fixed effect model with heterogeneous coefficients on the treatment.

Now let
\[
U_n = \left(\begin{array}{c} \beta_n\\ \mu_n\\1 \end{array}\right),
\quad  \tilde{U}_{t, a} = \left(\begin{array}{c}a \\ 1 \\ w_t \end{array}\right),\vspace{.2cm}
\]
and
\[
f_{t, a}(U_n) = \ldot{U_n}{\tilde{U}_{t, a}}.
\]
Then
\[
Y^{(a)}_{n, t} = f_{t, a}(U_n) + \varepsilon^{(a)}_{n, t},
\]
where $f_{t, a}$ is linear in $U_n$ (i.e., is H\"older continuous), and  $r = p+2$. 

Time-varying treatment coefficients can be easily accommodated in the same setting. Now suppose
\[
Y^{(a)}_{n, t} = \ldot{a}{\beta_{n, t}}+\mu_n + w_t + \varepsilon^{(a)}_{n, t}
\]
where $\beta_{n,t}=F_t\beta_n$, and $F_t$ is a $(p\times p)$ matrix of time-varying coefficients and the dimensions of the other components are unchanged.

Now let 
\[
U_n = \left(\begin{array}{c} \beta_n\\ \mu_n\\1 \end{array}\right),
\quad  \tilde{U}_{t, a} = \left(\begin{array}{c} \ldot{F_t}{a}  \\ 1 \\ w_t \end{array}\right),\vspace{.2cm}
\]
and
\[
f_{t, a}(U_n) = \ldot{U_n}{\tilde{U}_{t, a}}.
\]
Then again
\[
Y^{(a)}_{n, t} = f_{t, a}(U_n) + \varepsilon^{(a)}_{n, t},
\]
where $f_{t, a}$ is linear in $U_n$, and  $r = p+2$. 
\end{example}

\begin{example}[Interactive fixed effects model, \citeauthor{bai2009interactive}, \citeyear{bai2009interactive}]\label{example:interactive_FE}
Suppose
\[
Y^{(a)}_{n, t} = 
\ldot{a}{\beta} + \ldot{\mu_n}{w_t} + \varepsilon^{(a)}_{n, t},
\]
where $\mu_n$ and $v_t$ are factors of dimension $k$. 

Let 
\[
U_n = \left(\begin{array}{c} 1\\ \mu_n\end{array}\right),
\quad  \tilde{U}_{t, a} = \left(\begin{array}{c} \ldot{a}{\beta} \\ w_t \end{array}\right), 
\]
and
\[
f_{t, a}(U_n) = \ldot{U_n}{\tilde{U}_{t, a}},
\]
Then
\[
Y^{(a)}_{n, t} = f_{t, a}(U_n) + \varepsilon^{(a)}_{n, t},
\]
where $f_{t, a}$ is linear in $U_n$, and  $r = k+1$. 
\end{example}

\begin{example}[Tensor factor model, \cite{SI}]
Suppose 
\[
Y^{(a)}_{n, t} = 
\ldot{\mu_n}{w^{(a)}_t} + \varepsilon^{(a)}_{n, t},
\]
where $\mu_n, w^{(a)}_t$ are factors of dimension $k$.
One can verify the models in Examples \ref{example:two_way_FE} and \ref{example:interactive_FE} are special cases of the model above.
Here there are unit-specific heterogeneous coefficients on the treatment, and in addition the treatment can be time-varying.

Let 
\[
U_n = \mu_n,
\quad  \tilde{U}_{t, a} = w^{(a)}_t, 
\]
and
\[
f_{t, a}(U_n) = \ldot{U_n}{\tilde{U}_{t, a}},
\]
Then
\[
Y^{(a)}_{n, t} = f_{t, a}(U_n) + \varepsilon^{(a)}_{n, t},
\]
where $f_{t, a}$ is linear in $U_n$, and  $r = k$. 
\end{example}

\begin{example}[Dictionary basis expansion]
Consider 
\[
Y_{n, t}^{(a)}= \gamma_n(a, X_t) + \varepsilon^{(a)}_{n, t},
\]
where $X_t, a \in \Rb^p$, and $\gamma_n: \Rb^{2p} \to \Rb$ has the following dictionary representation
\[
\gamma_n(a, X_t) = \sum^L_{\ell = 1} \alpha_{n, \ell} b_{\ell}(a, X_t),
\]
where $b_{\ell}: \Rb^{2p} \to \Rb$ are dictionary basis functions, and $\alpha_{n, \ell} \in \Rb$, the corresponding linear coefficients. 
For example, $b_{\ell}$ could be a polynomial of $a$ and $X_t$.
Then, we can let
\[
U_n = \left(\begin{array}{c} \alpha_{n, 0}\\ \vdots \\ \alpha_{n, L} \end{array}\right),
\quad  \tilde{U}_{t, a} = \left(\begin{array}{c}b_{0}(a, X_t)\\ \vdots \\ b_{L}(a, X_t) \end{array}\right), \vspace{.2cm}
\] 
and
\[
f_{t, a}(U_n) = \ldot{U_n}{\tilde{U}_{t, a}}.
\]
Then
\[
Y^{(a)}_{n, t} = f_{t, a}(U_n) + \varepsilon^{(a)}_{n, t},
\]
where $f_{t, a}$ is linear in $U_n$, and $r = L$.
Note that in our model the dictionary basis functions $(b_{\ell})_{\ell \in [L]}$ and the covariates $X_t$ can be unobserved.
Further, our consistency results allow for $L$ to be increasing in $N, T$, as long as $L = o(\min(N, T))$.
\end{example}

\begin{example}[Binary choice]
Let $I_{\mathcal S}$ be the indicator function for set $\mathcal S$. 
Suppose
\[
Y_{n, t}^{(a)}=I_{[0,\infty)}\left(F\left(\ldot{a}{\beta} + \mu_n+w_t\right) - e_{n, t}^{(a)}\right),
\]
where $F: \Rb \to [0, 1]$ is a H\"{o}lder continuous function, and for every $(n, t, a)$, $e_{n, t}^{(a)}$ is an independent realization of a continuous random variable. 
Without loss of generality, we can assume that $e_{n, t}^{(a)}$ is uniformly distributed on $[0,1]$---if $e_{n, t}^{(a)}$ is not uniform on $[0,1]$, we can apply the probability integral transform, $f$, to both $F\left(\ldot{a}{\beta} + \mu_n + w_t\right)$ and $e_{n, t}^{(a)}$, where $f$ is the cumulative distribution function of $e_{n, t}^{(a)}$. 
Then, by Remark \ref{remark:non-linear_epsilon}, we can let
\[
U_n = \left(\begin{array}{c} 1\\ \mu_n\\1 \end{array}\right),
\quad  \tilde{U}_{t, a} = \left(\begin{array}{c}\ldot{a}{\beta} \\ 1 \\ w_t \end{array}\right).\vspace{.2cm}
\]
and
\[
f_{t, a}(U_n) = F\left(\ldot{U_n}{\tilde{U}_{t,a}}\right).
\]
Then
\[
Y^{(a)}_{n, t} = f_{t, a}(U_n) + \varepsilon^{(a)}_{n, t},
\]
where $f_{t, a}$ is H\"older continuous in $U_n$. 
Similar to the examples above, we can easily generalize this model to where we have $F\Big(\ldot{\mu_n}{w^{(a)}_{t}}\Big)$, i.e., we have unit-specific heterogeneous coefficients on treatments, and in addition, the treatments can be time-varying.
\end{example}

\subsection{Treatment Assignment Functions}\label{ssec:treatmentexample}
Without loss of generality, we will let $\Ac = \{0, 1\}$.
Below we show how various models for treatment assignment functions studied in the literature fit within our framework. 
In particular, our proposed DGP requires unconfoundedness conditional on $U_n$.
We formally establish how and when this condition holds for commonly studied treatment assignment functions.


We denote $h_{n, t}(\bU)$ as the $(n, t)$-th output of $h(\bU)$, i.e., the treatment $A_{n, t}$.

\begin{example}[Randomized trial]\label{example:RCT}
Consider a setup of a randomized trial where the $N$ units are assigned one of the two treatments at random, where the probability can differ across measurements.
Specifically, for all $n \in [N], t \in [T]$
\begin{align}
A_{n, t}
= \begin{cases}
1 &  \mbox{with probability}~p_t, \\
0 & \mbox{otherwise,} 
\end{cases}
\end{align}
independent of everything else.
Here, we can take
\[
h_{n, t}(\nu_{n, t}) = \mathbbm{1}\{v_{n,t}\leq p_t\},
\]
where $\nu_{n,t}$ is a random variable uniformly distributed in $[0, 1]$.
In this case, there is no confounding as the treatment assignment is not correlated with the outcomes, and so $h_{n, t}$ is not a function of $U_n$.

%
%
\end{example}

\begin{example}[Selection on (un)observables]\label{example:unobservables}
Suppose there are unobserved (or partially observed) covariates $U_n \in \Rb^q$ such that
\[
Y^{(a)}_{n, t} = f_{t, a}(U_n) + \varepsilon^{(a)}_{n, t}.
\]
Further, the treatment assignment for all $n \in [N], t \in [T]$ is given by
\begin{align}
A_{n, t}
= \begin{cases}
1 &  \mbox{with probability}~\sigma_{t}(U_n), \\
0 & \mbox{otherwise,} 
\end{cases}
\end{align}
where $\sigma_t$ is a function mapping to $[0, 1]$ (e.g., the logistic function).
Here $U_n$ is an unobserved confounder as it affects both the potential outcome $Y^{(a)}_{n, t}$, and is the input to the treatment assignment function $\sigma_{t}$.

Here, we can take
\[
h_{n, t}(U_n) = \mathbbm{1}\{v_{n,t}\leq\sigma_t(U_n)\},
\]
where $\nu_{n,t}$ is a random variable uniformly distributed in $[0, 1]$.
Our framework allows for treatment assignments where positivity does not hold, i.e., $\sigma_{t}(U_n)$ equals $0$ or $1$, by exploiting the smoothness of $f_{t, a}$ as given by Assumption \ref{assumption:holder.mech}.

%
\end{example}

\begin{example}[Regression discontinuity]\label{example:regdisc_score}
Suppose 
\[
Y^{(a)}_{n, t} = f_{t, a}(U_n) + \varepsilon^{(a)}_{n, t},
\]
as in Eq \eqref{eq:potential_outcomes_function}.
Further, suppose that treatment assignment for unit $n$  and measurement $t$ is a function of a score $X_{n, t}$.
In particular, treatment is given if the score $X_{n, t} \in \Rb^p$ is lower than some threshold $\theta_{n, t} \in \Rb^p$, i.e., 
\begin{align}
A_{n, t}
= \begin{cases}
1, &  \text{if} ~ X_{n, t} > \theta_{n, t} \\
0 &  \text{otherwise} 
\end{cases}
\end{align}
where 
\[
X_{n, t} = \ell_{n, t}(U_n),
\]
with $\ell_{n, t}: \Rb^q \to \Rb^p$. 
Here $U_n$ is an unobserved confounder as it affects both the potential outcome $Y^{(a)}_{n, t}$, and the score $X_{n, t}$, which in turn deterministically affects treatment assignment. 

Here, we can take 
\begin{align}
h_{n, t}(U_n) = \mathbbm{1}(\ell_{n, t}(U_n) > \theta_{n, t}).
\end{align}
As we show later, despite a regression discontinuity treatment assignment function, our framework allows for the estimation of treatment effects for units away from the threshold $\theta_{n, t}$ by exploiting the smoothness of $f_{t, a}$ as given by Assumption \ref{assumption:holder.mech}.

%
\end{example}

\begin{example}[Random utility model]\label{example:RUM}
Suppose 
\[
Y^{(a)}_{n, t} = f_{t, a}(U_n) + \varepsilon^{(a)}_{n, t},
\]
as in Eq \eqref{eq:potential_outcomes_function}.
Further, suppose that treatment assignment for unit $n$ and measurement $t$ is given as follows:
\begin{align}
A_{n, t}
= \begin{cases}
1, &  \text{if} ~ \ell_{t, 1}(U_n) - \ell_{t, 0}(U_n) + \nu_{n, t}  > \delta_{n, t} \\
0 &  \text{otherwise} 
\end{cases}
\end{align}
where $\ell_{t, 0}, \ell_{t, 1}: \Rb^q \to \Rb$, $\nu_{n, t}, \delta_{n, t} \in \Rb$.
If $\nu_{n, t}$ has a logistic distribution, then this recovers the Luce model (\cite{luce_model}). 

Here, we can simply take 
\begin{align}
h_{n, t}(U_n) = \mathbbm{1}(\ell_{t, 1}(U_n) - \ell_{t, 0}(U_n) + \nu_{n, t}  > \delta_{n, t}).
\end{align}


\end{example}

\begin{example}[Staggered adoption]\label{example:block}
Suppose we have a panel data model where $t$ corresponds to a time point and potential outcomes are given by
\[
Y^{(a)}_{n, t} = f_{t, a}(U_n) + \varepsilon^{(a)}_{n, t},
\]
as in Eq \eqref{eq:potential_outcomes_function}. 
Further, suppose that treatment assignment for unit $n$ and time $t$ is given as follows:
\begin{align}
A_{n, t}
= \begin{cases}
1 &  \mbox{if there exists\ \,} t' \le t,  \mbox{\ such that\ \,} U_n > \theta_{n, t'} \\
0 & \mbox{otherwise,} 
\end{cases}
\end{align} 
where $\theta_{n, t'} \in \Rb^q$.
That is, unit $n$ receives treatment $A_{n,t} = 1$ for time period $t$ if there existed a time point $t' \le t$ such that $U_n$ is less than the threshold $\theta_{n, t'}$, which is both unit and time specific. 
Here assignment of intervention $1$ is an absorbing state.
It is easy to see that such an assignment scheme leads to a staggered adoption observation pattern.

Here, we can take 
\begin{align}
h_{n, t}(U_n) = \mathbbm{1}(U_n > \bar{\theta}_{n, t}),
\end{align}
where $\bar{\theta}_{n, t} = \min\{\theta_{n, 1}, \ldots, \theta_{n, t}\}$.

\end{example}

\section{Identification and Estimation of Treatment Effects}\label{sec:outcomes_factor_structure}

{\bf Causal parameters of interest.}
We restrict our attention to the binary treatment setting where for all $t \in [T]$, we let $\Ac = \{0, 1\}$.
Our analysis easily extends for any finite $\Ac$.
We focus on the estimation of the average treatment effect for a given measurement $t^*$, and for a subset of units $\Mc \subset [N]$, with $|\Mc| = M$: 
\begin{align}
\Ex[\mathsf{ATE}_{\Mc} \mid \bU] = \frac{1}{M}\sum_{n \in \Mc} \Ex\Big[\Big(Y^{(1)}_{n, t^*} - Y^{(0)}_{n, t^*} \Big) \mid \bU\Big],
\end{align}
where expectations are taken over the distribution $\varepsilon^{(a)}_{n, t^*}$.
We note our target causal parameter is defined conditional on the unobserved confounders $\bU$.

Let 
\begin{align}
\Ic^{(a)} = \{n \in [N] : A_{n, t^*} = a\}, \quad N_a = |\Ic^{(a)}|.
\end{align}
That is, $\Ic^{(a)}$ is the set of units that received intervention $a$ for measurement $t^*$, and $N_a$ is the number of units in that set.
For different subsets $\Mc$, $\mathsf{ATE}_{\Mc}$ nests a variety of causal parameters of interest:
\begin{itemize}
\item If $\Mc = \Ic^{(1)}$ (all the units that were treated during measurement $t^*$), then $\mathsf{ATE}_{\Mc}$ corresponds to the {\em average treatment effect on the treated}, which we denote as $\mathsf{ATT}$.
\item If $\Mc = \Ic^{(0)}$ (all the units that were untreated during measurement $t^*$), then $\mathsf{ATE}_{\Mc}$ corresponds to the {\em average treatment effect on the  untreated}, which we denote as $\mathsf{ATU}$.
\item If $\Mc = [N]$ , then $\mathsf{ATE}_{\Mc}$ corresponds to the {\em average treatment effect}, which we denote as $\mathsf{ATE}$.
\end{itemize}
For concreteness, we focus our estimation results on $\mathsf{ATT}, \mathsf{ATU}$, and $\mathsf{ATE}$. However, our results easily extend to any set of units $\Mc \subset [N]$.

\subsection{Identification}
We now show how the model for treatment assignment and potential outcomes, summarized in Section \ref{sec:model} leads to a novel identification argument for $\mathsf{ATE}_{\Mc}$.
Motivated by Proposition \ref{prop:factor_model_approx}, we define the linear factor model approximation error to $f_{t, a}(U_n)$ as follows.
\begin{definition}[Linear factor model approximation]\label{def:linear_factor_approx}
For $r \in \Nb$, let $\{\lambda_n \}_{n \in [N]} \cup  \{ \rho_{t, a} \}_{t \in [T], a \in \Ac}$, with $\lambda_n, \rho_{t, a} \in \Rb^r,$ be (one of) the linear factor model approximations of $f_{t, a}(U_n)$ that minimizes $\Delta_E$ where, 
\begin{align}
\Delta_E = \max_{n \in [N], t \in [T], a \in \Ac}| \eta^{(a)}_{n, t} |, \quad \text{and } \eta^{(a)}_{n, t} = f_{t, a}(U_n) - \ldot{\lambda_{n}}{\rho_{t, a}}.
\end{align}
\end{definition}
Recall that if $f_{t, a}$ is H\"older continuous, then Proposition \ref{prop:factor_model_approx} implies that both $r$ and $\Delta_E$ can be simultaneously controlled.

We define two subsets of $\Mc$: for $a \in \{0, 1\}$,
\begin{align}
\Mc^{(a)} = \{ n \in \Mc : A_{n, t^*} = a \},
\quad M_a = |\Mc^{(a)}|.
\end{align}
Note that $\Mc^{(a)} \subset \Ic^{(a)}$.
We are now equipped to define the key assumption we require for identification of the causal parameter of interest.
\begin{assumption}[Linear span inclusion]\label{assumption:linear_span_inclusion}
For $a \in \{0, 1\}$, let $\lambda_{\Mc^{(1 - a)}} = \sum_{n \in \Mc^{(1 - a)}} \lambda_n$.
We assume there exists linear weights $\beta^{(a)} \in \Rb^{N_{a}}$ such that,
\begin{align}\label{equation:weights}
\lambda_{\Mc^{(1 - a)}} = \sum_{n \in \Ic^{(a)}} \beta^{(a)}_n \lambda_n.
\end{align}
That is, $\lambda_{\Mc^{(1 - a)}}$ lies in the linear span of $\{\lambda_n\}_{n \in \Ic^{(a)}}$.
In settings where there are multiple weights that satisfy condition \eqref{equation:weights}, we define $\beta^{(a)}$ to be the unique one with minimum $\ell_2$-norm.
\end{assumption}
This assumption implicitly adds a restriction on the treatment assignment.
For example, Assumption \ref{assumption:linear_span_inclusion} does not allow for a treatment assignment mechanism such that the latent factors associated with the units in $\Ic^{(a)}$ and $\Mc^{(1 - a)}$ live in orthogonal spaces.
Hence the assignment mechanism needs to be diverse enough, so that the latent factors associated with the units in different treatments are linearly expressible in terms of each other.
We only require the weaker condition that this linear span inclusion holds for the {\em sum} of the unit latent factors associated with $\Mc^{(1 - a)}$, as opposed to it holding for {\em each} latent factor $\lambda_n$ for $n \in \Mc^{(1 - a)}$. 

\begin{theorem}[Identification]\label{thm:ate_idenfication}
Let Assumptions \ref{assumption:DGP}, \ref{assumption:holder.mech} and \ref{assumption:linear_span_inclusion} hold.
Then, given $\beta^{(a)}$ for $a \in \{0, 1\}$,
\begin{align}\label{eq:ate_identification}
\sum_{n \in \Mc} \Ex[Y^{(a)}_{n, t^*} \mid \bU]
&= \sum_{n \in \Mc^{(a)}} \Ex[Y_{n, t^*} \mid \bA, \bU] 
+ \sum_{n \in\Ic^{(a)}} \beta^{(a)}_n \Ex\left[Y_{n, t^*} \mid \bA, \bU\right]
- \sum_{n \in \Ic^{(a)}} \beta^{(a)}_n \eta^{(a)}_{n, t^*}
+ \sum_{n \in \Mc^{(1-a)}} \eta^{(a)}_{n, t^*},\quad
\end{align}
where expectations are taken over the distribution of $\varepsilon^{(a)}_{n, t^*}$.
\end{theorem}
We note that an explicit representation of $\beta^{(a)}$ in terms of the observed data is given in \eqref{eq:PCR_remark_1} below, and we establish $\hbeta^{(a)}$ as given in \eqref{eq:PCR_step2} is a consistent estimator for it in Proposition \ref{prop:pcr_est}.
Next, we establish how Theorem \ref{thm:ate_idenfication} helps establish identification of our causal parameter of interest.

\begin{corollary}[Identification]\label{cor:ate_identification}
Let 
\begin{align}
\mathsf{Observed}_{a} = 
\frac{1}{M}\left(\sum_{n \in \Mc^{(a)}} \Ex[Y_{n, t^*} \mid \bA, \bU] 
+ \sum_{n \in \Ic^{(a)}} \beta^{(a)}_n \Ex\left[ Y_{{n, t^*}} \mid \bA, \bU \right]\right),
\end{align}
and
\begin{align}
\mathsf{Observed} =
\mathsf{Observed}_{1} - \mathsf{Observed}_{0}.
\end{align}
Then, under the conditions of Theorem \ref{thm:ate_idenfication},
\begin{align}\label{eq:ate_identification_full}
\Big| \Ex[\mathsf{ATE}_{\mathcal{M}} \mid \bU] - \mathsf{Observed} \Big| 
\le \Delta_E \left(1 + \frac{\| \beta^{(0)}\|_1 + \| \beta^{(1)}\|_1}{M} \right).
\end{align}
\end{corollary}


\subsection{Estimator}
The identification result in Corollary \ref{cor:ate_identification} suggests an estimator of the form 
\begin{align}\label{eq:ATE_synthetic_estiamtor}
\widehat{\mathsf{ATE}}_{\Mc}
= \frac{1}{M} \left(\sum_{n \in \Mc^{(1)}} Y_{n, t^*} 
+ \sum_{n \in \Ic^{(1)}} \hbeta^{(1)}_n Y_{n, t^*} \right)  
- \frac{1}{M} \left(\sum_{n \in \Mc^{(0)}} Y_{n, t^*} 
+  \sum_{n \in \Ic^{(0)}} \hbeta^{(0)}_n Y_{n, t^*} \right),
\end{align}
where $\hbeta^{(1)}_j$ and $\hbeta^{(0)}_j$ are estimates of $\beta^{(1)}_j$ and $\beta^{(0)}_j$, respectively. 
That is, $\widehat{\mathsf{ATE}}_{\Mc}$ imputes the sums of the unobserved potential outcomes with and without treatment by $\sum_{n \in \Ic^{(1)}} \hbeta^{(1)}_n Y_{n, t^*}$ and $\sum_{n \in \Ic^{(0)}} \hbeta^{(0)}_n Y_{n, t^*}$, respectively.

Below, we provide sufficient conditions on any estimator $\hbeta^{(a)}$ of $\beta^{(a)}$, which establish the finite-sample consistency of $\widehat{\mathsf{ATE}}_{\Mc}$.
Hence, we denote 
\begin{align}
\Delta_{\beta^{(a)}} = \hbeta^{(a)} - \beta^{(a)}.
\end{align}
In Section \ref{sec:PCR}, we provide explicit conditions for consistency and normality when the estimator for $\hbeta^{(a)}$ is PCR.

\subsection{Finite-Sample Consistency}\label{sec:abstract_finite_sample}
To establish consistency, we make the (mild) assumption that $\varepsilon^{(a)}_{n, t}$ has a sub-Gaussian distribution.
\begin{assumption}[Sub-Gaussian potential outcomes]\label{assumption:sub-gaussian_noise}
For all $(n, t, a)$, $\varepsilon^{(a)}_{n, t} \mid \bU$ is a sub-Gaussian random variable with standard deviation $\sigma^{(a)}_{n, t}$.
Let $\sigma_\max = \max_{n \in [N], t \in [T], a \in \{0, 1\}} \sigma^{(a)}_{n, t}$, and assume $\sigma_\max < C$.
\end{assumption}

\begin{proposition}[Conditions for consistency for any linear estimator]\label{prop:ate_estimation}
Let Assumptions \ref{assumption:DGP}, \ref{assumption:holder.mech}, \ref{assumption:linear_span_inclusion}, and \ref{assumption:sub-gaussian_noise} hold.
Let $Y_{\Ic^{(a)}} = [Y_{n, t^*}]_{n \in \Ic^{(a)}}, \varepsilon_{\Ic^{(a)}} \coloneqq [\varepsilon^{(a)}_{n, t^*}]_{n \in \Ic^{(a)}}$.
Then,
\begin{align}
\widehat{\mathsf{ATE}}_{\Mc} &- \Ex[\mathsf{ATE}_{\Mc} \mid \bU] \le \textsf{Bias} + \textsf{Variance}
\end{align}
where
\begin{align}
\textsf{Bias} &= \Delta_E \left(1 + \frac{\| \beta^{(0)}\|_1 + \| \beta^{(1)}\|_1}{M} \right) + O_p\left( \sum_{a \in \{0, 1\}} \frac{\sigma_\max \|\Delta_{\beta^{(a)}}\|_2 +  \ldot{\Delta_{\beta^{(a)}}  }{\Ex\left[ Y_{\Ic^{(a)}} \right]}}{M} \right) \label{eq:prop_ate_est_1}
\\ \textsf{Variance} &=  O_p\left( \sum_{a \in \{0, 1\}} \frac{\sigma_\max \left(\sqrt{M_a} +  \|\beta^{(a)} \|_2 \right)}{M} \right) \label{eq:prop_ate_est_2}
\end{align}
\end{proposition}


\section{Estimation results using Principal Component Regression (PCR)}\label{sec:PCR}%

In Section \ref{sec:ate_pcr} below, we provide explicit bounds on $\Delta_{\beta^{(a)}}$ for the case when PCR is used to estimate the coefficients $\hbeta^{(a)}$.
Theorem \ref{thm:consistency_summary} collects sufficient conditions for consistency of $\widehat{\mathsf{ATE}}_{\Mc}$.

\subsection{Estimating Linear Weights via PCR}\label{sec:ate_pcr}
We now show how to estimate $\hbeta^{(a)}$ via PCR, and subsequently control $\Delta_{\beta^{(a)}}$ for the case when there are many measurements of all units under a common set of interventions. 

\noindent
{\bf Necessary notation to define PCR.}
We introduce additional notation that we will use to discuss PCR estimation of $\hbeta^{(a)}$. 
Let $\bar{\Tc} \subset [T]$ be defined as follows:
\begin{align}
\bar{\Tc} = \{ t \in [T]: A_{n, t} = A_{n', t}\,\, \forall n, n' \in [N]\},
\quad \bar{T} = |\bar{\Tc}|.
\end{align}
That is, $\bar{\Tc}$ is the of measurements for which all units are seen under the same intervention.
Let $a_t$ be the common treatment value for $t \in \bar{\Tc}$.
For $a \in \{0, 1\}$, define
\begin{align}
\bY &= \left[\sum_{n \in \Mc^{(1 - a)}}Y_{n, t}\right]_{t \in \bar{\Tc}} \in \Rb^{\bar{T}},
\\ \bZ &= \Big[ Y_{n, t} \Big]_{t \in \bar{\Tc}, j \in \Ic^{(a)}} \in \Rb^{\bar{T} \times N_a},
\\ \bX &= \Big[ \Ex[Y_{n, t}] \Big]_{t \in \bar{\Tc}, j \in \Ic^{(a)}} \in \Rb^{\bar{T} \times N_a},
\\ \bX^{\text{lr}} &= \Big[\ldot{\lambda_{n}}{\rho_{t, a_t}}\Big]_{t \in \bar{\Tc}, n \in \Ic^{(a)}} \in \Rb^{\bar{T} \times N_a},
\end{align}
where to reduce notational burden we suppress dependence on $a$ in the notation for $\bY$, $\bZ$, and $\bX$.
$\bY$ is a vector of summed outcomes of the units in $\Mc^{(1 - a)}$ for the measurements in $\bar{\Tc}$,
$\bZ$ is a matrix of outcomes for the units in $\Ic^{(a)}$, and measurements in $\bar{\Tc}$, and $\bX$ is defined analogously to $\bZ$, but with respect to the expected observed outcomes.
$\bX^{\text{lr}}$ is the low-rank approximation of $\bX$; note $\bX - \bX^{\text{lr}} = [\eta^{(a_t)}_{n, t}]_{t \in \bar{\Tc}, n \in \Ic^{(a)}}$.

\noindent
{\bf PCR estimator for $\hbeta^{(a)}$.}
Define the singular value decomposition (SVD) of $\bZ$ as 
\begin{align}
\bZ = \sum^{\min(\bar{T}, N_a)}_{\ell = 1} \hat{s}_\ell \hat{u}_{\ell} \hat{v}^T_{\ell},
\end{align}
where $\hat{s}_\ell, \hat{u}_\ell, \hat{v}_\ell$ refer to the $\ell$-th singular value, left singular vector, and right singular vector, respectively.
For any SVD, we order the singular values by decreasing magnitude.
Given hyper-parameter $k \in [\min(\bar{T}, N_a)]$, we define $\widehat{\bX}^{\text{lr}}$ as follows:
\begin{align}\label{eq:PCR_step1}
\widehat{\bX}^{\text{lr}} = \sum^{k}_{\ell = 1} \hat{s}_\ell \hat{u}_{\ell} \hat{v}^T_{\ell}.
\end{align}
That is, $\widehat{\bX}^{\text{lr}}$ is a low-rank approximation of $\bZ$.
$\hbeta^{(a)}$ is then estimated by simply doing ordinary least squares (OLS) on $\bY$ and $\widehat{\bX}^{\text{lr}}$ as follows:
\begin{align}\label{eq:PCR_step2}
\hbeta^{(a)} = \Big(\widehat{\bX}^{\text{lr}})^{+} \bY.
\end{align}
Here $\Big(\widehat{\bX}^{\text{lr}})^{+}$ denotes the Moore-Penrose pseudoinverse of $\widehat{\bX}^{\text{lr}}$ defined as
\begin{align}
\Big(\widehat{\bX}^{\text{lr}})^{+} = \left( \sum^k_{\ell = 1}\frac{\hat{v}_{\ell} \hat{u}^T_{\ell}}{\hat{s}_\ell}  \right).
\end{align}
That is, PCR can be seen as doing ordinary least squares (OLS) on the best $k$-rank approximation of $\bZ$, given by $\Big(\widehat{\bX}^{\text{lr}})^{+}.$ 
If the ordinary least squares problem has multiple solutions, it is well-known that $\hbeta^{(a)}$ in equation \eqref{eq:PCR_step2} is the minimum $\ell_2$-norm solution.
We can then use $\hbeta^{(a)}$, to estimate $\widehat{\mathsf{ATE}}_{\Mc}$ as shown in \eqref{eq:ATE_synthetic_estiamtor}.

\noindent
{\bf Interpreting PCR.}
Using Assumption \ref{assumption:holder.mech} and Definition \ref{def:linear_factor_approx}, we have that for all $n \in [N], t \in \bar{\Tc}$, 
\begin{align}\label{eq:PCR_remark_0}
Y_{n, t} 
= \ldot{\lambda_{n}}{\rho_{t, a_t}} 
+  \eta^{(a_t)}_{n, t} 
+ \varepsilon^{(a_t)}_{n, t}.
\end{align}
Hence, using \eqref{eq:PCR_remark_0} and the definitions of $\bY, \bZ, \bX, \bX^{\text{lr}}$, we have
\begin{align}
Y_{t}  
&= \sum_{n \in \Mc^{(1 - a)}} \left(\ldot{\lambda_{n}}{\rho_{t, a_t}} +  \eta^{(a_t)}_{n, t} + \varepsilon^{(a_t)}_{n, t} \right) 
\\ \bX^{\text{lr}}_{t, n} 
&= \ldot{\lambda_n}{\rho_{t, a_t}}  
\\ \bX_{t, n} 
&= \ldot{\lambda_n}{\rho_{t, a_t}} + \eta^{(a_t)}_{n, t} 
\\ \bZ_{t, n} 
&= \ldot{\lambda_n}{\rho_{t, a_t}} + \eta^{(a_t)}_{n, t} +\varepsilon^{(a_t)}_{n, t} 
\end{align}
By Assumption \ref{assumption:linear_span_inclusion}, we have 
$\lambda_{\Mc^{(1 - a)}}  = \sum_{n \in \Ic^{(a)}} \beta^{(a)}_{n} \lambda_n$.
Hence, we can write
\begin{align}
\bZ &= \bX + \bH \label{eq:PCR_remark_3}
\\ \bX &= \bX^{\text{lr}} + \bE^{\text{lr}} \label{eq:PCR_remark_2}
\\ \bY &= \bX^{\text{lr}}\beta^{(a)}  
+ \phi^{\text{lr}} + \bar{\varepsilon} \label{eq:PCR_remark_1}
\end{align}
where 
$\phi^{\text{lr}} = \left[\sum_{n \in \Mc^{(1 - a)}} \eta^{(a_t)}_{n, t} \right]_{t \in \bar{\Tc}} \ $, 
$\bar{\varepsilon} = \left[\sum_{n \in \Mc^{(1 - a)}} \varepsilon^{(a_t)}_{n, t} \right]_{t \in \bar{\Tc}} \ $, 
$\bE^{\text{lr}} = [\eta^{(a_t)}_{n, t}]_{t \in \bar{\Tc}, n \in \Ic^{(a)}} \ $, 
$\bH = [\varepsilon^{(a_t)}_{n, t}]_{t \in \bar{\Tc}, n \in \Ic^{(a)}}$.
Thus we have reduced our problem of estimating $\beta^{(a)}$ to that of linear regression where:
(i) the covariates are noisily observed (i.e., error-in-variables regression), i.e. \eqref{eq:PCR_remark_3} holds;
(ii) the noiseless covariate matrix has an approximate low-rank representation, i.e. \eqref{eq:PCR_remark_2} holds;
(iii) an approximate linear model holds between the approximate low-rank representation of the noiseless covariates and the response variable, i.e. \eqref{eq:PCR_remark_1} holds.

Hence, we can interpret PCR as follows:
(1) the first step of doing PCA given in \eqref{eq:PCR_step1} creates an estimate of the approximate low-rank approximation $\bX^{\text{lr}}$; 
(2) the second step of doing OLS given in \eqref{eq:PCR_step2} creates an estimate of $\beta^{(a)}$ by regressing $\bY$ on $\widehat{\bX}^{\text{lr}}$, which is motivated by \eqref{eq:PCR_remark_1}.

The novel technical challenge in analyzing this setting is that there are four sources of error: the noise on the covariates given by $\bH$, the low-rank approximation error given by $\bE^{\text{lr}}$, the linear model approximation error given by $\phi^{\text{lr}}$, and the error on the response given by $\bar{\varepsilon}$.

\subsection{Additional Assumptions for Estimation Results with PCR}
We make the following additional assumptions to state our consistency results. 
Note by Assumption \ref{assumption:holder.mech} and Proposition \ref{prop:factor_model_approx}, for all $\delta > 0$,
\[
\text{rank}(\bX^{\text{lr}}) := \bar{r} \le r \le \bar{C} \cdot \delta^{-q}, \quad
\|\bX - \bX^{\text{lr}} \|_{\infty} = \Delta_E \le C_H \cdot \delta^S.
\]

\begin{remark}
Previewing our consistency results, we will pick $\delta = \left( \frac{1}{(\min(N_0, N_1, \bar{T})}\right)^{\frac{1}{2S}}$.
Then, $r \le \bar{C} \min(N_0, N_1, \bar{T})^{\frac{q}{2S}}$ and $\Delta_E \le C_H \left(\frac{1}{(\min(N_0, N_1, \bar{T})}\right)^{\frac{1}{2}}.$
Hence, if $q < 2S$, then as $\min(N_0, N_1, \bar{T}) \to \infty$, $r \ll \min(N_0, N_1, \bar{T})$ and $\Delta_E = o(1)$.
\end{remark}


\begin{assumption}[Well-balanced spectra.]\label{assumption:pcr_well_balanced}
Given the SVD of $\bX^{\text{lr}} = \sum^{\bar{r}}_{\ell = 1} s_\ell u_{\ell} v^T_{\ell},$ we assume
\begin{align}
s_{\bar{r}} \ge C \sqrt{\frac{ \bar{T} N_a }{\bar{r}}}.   
\end{align}
\end{assumption}
An interpretation of Assumption \ref{assumption:pcr_well_balanced} is as follows.
Suppose that each entry of $\bX^{\text{lr}} \ge c > 0$, i.e. is bounded below by an absolute constant $c$.
Then since $\bX^{\text{lr}} \in \Rb^{\bar{T} \times N_a}$ we have that $\sum^{\bar{r}}_{\ell = 1} s^2_\ell = \| \bX^{\text{lr}} \|^2_F \ge C \bar{T} N_a$.
If all the singular values of $\bX^{\text{lr}}$ are of the same order of magnitude, i.e., $\frac{s_{\bar{r}}}{s_1} \ge C$, this immediately implies that $s_r \ge C \sqrt{\frac{ \bar{T} N_a }{\bar{r}}}$.

\begin{assumption}[Subspace inclusion.]\label{assumption:pcr_subspace}
For intervention $a \in \{0, 1\}$, $\Big[\ldot{\lambda_n}{\rho_{t^*, a}}\Big]_{n \in \Ic^{(a)}}$ lies in the rowspace of $\bX^{\text{lr}} = \Big[\ldot{\lambda_{n}}{\rho_{t, a_t}} \Big]_{t \in \bar{\Tc}, n \in \Ic^{(a)}}$.
\end{assumption}
Note a sufficient condition for Assumption 
\ref{assumption:pcr_subspace} is for $a \in \{0, 1\}$ 
\begin{align}
\rho_{t^*, a} \in \text{span}\{\rho_{t, a_t} \}_{t \in \bar{\Tc}}.
\end{align}
Hence, intuitively, we require that the target measurement $t^*$ for which we wish to compute $\mathsf{ATE}_{\Mc}$, the latent factors $\rho_{t^*, 0}, \rho_{t^*, 1}$, are linearly expressible in terms of the latent factors $\{\rho_{t, a_t} \}_{t \in \bar{\Tc}}$ corresponding to the measurements under which all units are under a common intervention.
This is the key condition that lets us {\em generalize} from the measurements $\bar{T}$ we learn on to the measurement $t^*$ we make counterfactual predictions on.

Below, we provide exact conditions if the linear estimator is PCR, the appropriateness of which was motivated in Section \ref{sec:ate_pcr}.

\subsection{Finite-sample Consistency using PCR}

\begin{theorem}[ATT, ATU, ATE consistency using PCR]\label{thm:consistency_summary}
Let Assumptions \ref{assumption:DGP}, 
\ref{assumption:holder.mech} , \ref{assumption:linear_span_inclusion}, \ref{assumption:sub-gaussian_noise}, \ref{assumption:pcr_well_balanced}, and \ref{assumption:pcr_subspace} hold.
Let $\hbeta^{(a)}$ be estimated via PCR as in \eqref{eq:PCR_step1} and \eqref{eq:PCR_step2}.
Assume the following additional conditions hold.
\begin{enumerate}
\item Correct rank estimation for PCR: $k$ in \eqref{eq:PCR_step1} is such that $k = \bar{r}$.
\item Smooth outcome model: Let $\alpha = S / q > 0$, where recall $S$ is smoothness parameter of $f_{t, a}(U_n)$ and $q$ is the latent dimension of $U_n$.  Assume $\alpha > 1.$
\item Disperse weights: For $a \in \{0, 1\}$, assume $\|\beta^{(a)}\|_2 = O\left( \frac{M_{1 - a}}{N^{w}_a} \right)$, where $\frac{1}{2\alpha} < w \le \frac{1}{2}$.
\item Growing common measurements, units: $\min(N_0, N_1, \bar{T}) \to \infty.$
\end{enumerate}
Then we have the following consistency results:

\begin{itemize}
\item {\bf $\mathsf{ATT}$ consistency:}
If $\frac{N_0^{1 - w}}{\bar{T}^{1 - \frac{1}{2 \alpha}}} = o(1)$, we have,
\begin{align}
\widehat{\mathsf{ATT}}  - \Ex[\mathsf{ATT} \mid \bU] &= o_p(1).
\end{align}
Further, we can take 
\begin{align}
r \le \bar{C} \min(N_0, \bar{T})^{\frac{1}{2\alpha}}, \quad \Delta_E \le C \min(N_0, \bar{T})^{- \frac{1}{2}}.
\end{align}

\item {\bf $\mathsf{ATU}$ consistency:}
If $\frac{N_1^{1 - w}}{\bar{T}^{1 - \frac{1}{2 \alpha}}} = o(1)$, we have,
\begin{align}
\widehat{\mathsf{ATU}}  - \Ex[\mathsf{ATU} \mid \bU] &= o_p(1).
\end{align}
Further, we can take 
\begin{align}
r \le \bar{C} \min(N_1, \bar{T})^{\frac{1}{2\alpha}}, \quad \Delta_E \le C \min(N_1, \bar{T})^{- \frac{1}{2}}.
\end{align}

\item {\bf $\mathsf{ATE}$ consistency:}
If $N_0, N_1 = \Theta(N)$, and $\frac{N^{1 - w}}{\bar{T}^{1 - \frac{1}{2 \alpha}}} = o(1)$, we have,
\begin{align}
\widehat{\mathsf{ATE}}  - \Ex[\mathsf{ATE} \mid \bU] &= o_p(1).
\end{align}
Further, we can take 
\begin{align}
r \le \bar{C} \min(N, \bar{T})^{\frac{1}{2\alpha}}, \quad \Delta_E \le C \min(N, \bar{T})^{- \frac{1}{2}}.
\end{align}

\end{itemize}
\end{theorem}
Theorem \ref{thm:consistency_summary} establishes exact conditions such that PCR is a consistent estimate for $\widehat{\mathsf{ATT}}, \widehat{\mathsf{ATU}}$, and $\widehat{\mathsf{ATE}}$, which are quantified by: the smoothness of $f_{t, a}$; the dimension of $U_n$; the number of common measurements $\bar{T}$; the number of units undergoing interventions $a \in \{0, 1\}$ given by $N_0, N_1$; and the magnitude of the linear weights $\beta^{(a)}$.
We recall from Section \ref{sec:latent_strucutre_outcomes} that if $f_{t, a}$ is an analytic function, then we can take $S$ to be an arbitrarily large integer, i.e., it is  H\"older continuous for all $S \in \Nb$.

Below we provide a natural sufficient condition for which the disperse weights condition holds.
\begin{proposition}\label{prop:beta_norm}
Assume for every set $\Ic \subset [N]$ where $|\Ic| = N^{\theta}$, with $0 < \theta < 1 -  \frac{3}{2\alpha}$, there exists a subset $\tilde{\Ic}$, where $|\tilde{\Ic}| = r$ and $\text{span}\{ \lambda_n\}_{j \in \tilde{\Ic}} = \Rb^r$.
Assume $r \le \bar{C}\min(N_a, \bar{T})^{\frac{1}{2\alpha}}$.
Then the minimum $\ell_2$-norm $\beta^{(a)}$ is such that $\|\beta^{(a)}\|_2 =o\left(\frac{M_{1 - a}}{N_{a}^{{\frac{1}{2\alpha}}}}\right)$.
That is, the property, $\frac{1}{2 \alpha} < w$, in Condition 3 of Theorem \ref{thm:consistency_summary} holds.
\end{proposition}
Proposition \ref{prop:beta_norm} establishes that if for any given set of units of sufficient size, their associated latent factors space the entire space $\Rb^r$, then the disperse weights condition must hold.



\newpage
\bibliographystyle{apalike}
\bibliography{bib}

\newpage
\begin{appendix}
\section{ATE: Identification Proofs}
\subsection{Proof of Theorem \ref{thm:ate_idenfication}}
From Assumption \ref{assumption:DGP}, we have that for $a \in \{0, 1\},$
\begin{align}
\sum_{n \in \Mc} \Ex[Y^{(a)}_{n, t^*} \mid U_n]
&= \sum_{n \in \Mc^{(a)}} \Ex[Y^{(a)}_{n, t^*} \mid U_n] + \sum_{n \in \Mc^{(1 - a)}} \Ex[Y^{(a)}_{n, t^*} \mid U_n]\\
&= \sum_{n \in \Mc^{(a)}} \Ex[Y^{(a)}_{n, t^*} \mid \bA, \bU] + \sum_{n \in \Mc^{(1 - a)}} \Ex[Y^{(a)}_{n, t^*} \mid \bU]
\\&= \sum_{n \in \Mc^{(a)}} \Ex[Y_{n, t^*} \mid \bA, \bU] + \sum_{n \in \Mc^{(1 - a)}} \Ex[Y^{(a)}_{n, t^*} \mid \bU]. \label{eq:ate_id_1}
\end{align}
What remains to be tackled is the second term in \eqref{eq:ate_id_1}.
Assumptions \ref{assumption:DGP} and \ref{assumption:holder.mech}, and
Definition \ref{def:linear_factor_approx} implies $Y^{(a)}_{n, t^*} = \ldot{\lambda_n}{\rho_{t^*, a}} + \eta^{(a)}_{n, t^*} + \varepsilon^{(a)}_{n, t^*}$, where $\Ex[\varepsilon^{(a)}_{n, t^*} \mid \bU] = 0$.
Hence, from Assumption \ref{assumption:linear_span_inclusion},
\begin{align}
\sum_{n \in \Mc^{(1 - a)}} \Ex[Y^{(a)}_{n, t^*} \mid \bU]
&= \sum_{n \in \Mc^{(1 - a)}} 
\left(\ldot{\lambda_n}{\rho_{t^*, a}} + \eta^{(a)}_{n, t^*}\right)
\nonumber\\ &= \ldot{\lambda_{\Mc^{(1 - a)}}}{\rho_{t^*, a}}  + \sum_{n \in \Mc^{(1-a)}} \eta^{(a)}_{n, t^*}\nonumber\\
&= \sum_{n \in \Ic^{(a)}} \beta^{(a)}_n  \ldot{\lambda_n}{\rho_{t^*, a}}  + \sum_{n \in \Mc^{(1-a)}} \eta^{(a)}_{n, t^*}
\label{eq:ate_id_2}
\end{align}
where in the last line we have used the definition of $\lambda_{\Mc^{(1 - a)}}$.

In addition,
\begin{align}
\sum_{n \in \Ic^{(a)}} \beta^{(a)}_n \ldot{  \lambda_n}{\rho_{t^*, a}}
%
&= 
\sum_{n \in\Ic^{(a)}} \beta^{(a)}_n \Ex\left[Y^{(a)}_{n, t^*} \mid \bU\right]
- \sum_{n \in \Ic^{(a)}} \beta^{(a)}_n \eta^{(a)}_{n, t^*}\nonumber\\
&=\sum_{n \in\Ic^{(a)}} \beta^{(a)}_n \Ex\left[Y_{n, t^*} \mid \bA, \bU\right]
- \sum_{n \in \Ic^{(a)}} \beta^{(a)}_n \eta^{(a)}_{n, t^*}.
\label{eq:ate_id_3}
\end{align}

Combining \eqref{eq:ate_id_1}, \eqref{eq:ate_id_2}, and \eqref{eq:ate_id_3}, we conclude the proof. 

\subsection{Proof of Corollary \ref{cor:ate_identification}.}
Using Theorem \ref{thm:ate_idenfication}, we have 
\begin{align}
\Big| \Ex[\mathsf{ATE}_{\mathcal{M}} \mid \bU] - \mathsf{Observed} \Big| 
&\le  \frac{1}{M} \left(\left| \sum_{j \in \Mc^{(0)}} \eta^{(1)}_{n, t^*}\right|
+ \left| \sum_{j \in \Ic^{(1)}} \beta^{(1)}_j \eta^{(1)}_{n, t^*}\right|
+ \left| \sum_{j \in \Mc^{(1)}} \eta^{(0)}_{n, t^*}\right|
+ \left| \sum_{j \in \Ic^{(0)}} \beta^{(0)}_j \eta^{(0)}_{n, t^*}\right|\right) \label{eq:ate_identification_cor_1}
\\&\le  \Delta_E \left(1 + \frac{\| \beta^{(0)}\|_1 + \| \beta^{(1)}\|_1}{M} \right). \label{eq:ate_identification_cor_2}
\end{align}


\section{ATE: Estimation Proofs}

\subsection{Proof of Proposition \ref{prop:ate_estimation}.}

We recall notation required for the proofs of this section.
Let $\varepsilon_{\Ic^{(a)}} \coloneqq [\varepsilon^{(a)}_{n, t^*}]_{n \in \Ic^{(a)}}$, $Y_{\Ic^{(a)}} \coloneqq [Y_{n, t^*}]_{n \in \Ic^{(a)}}$.
%

From Corollary \ref{cor:ate_identification} and the definition of a linear estimator in \eqref{eq:ATE_synthetic_estiamtor}, we have that
\begin{align}
|&\widehat{\mathsf{ATE}}_{\Mc}-\Ex\left[\mathsf{ATE}_{\Mc} \mid \bU \right] | \\
&\le  \Delta_E \left(1 + \frac{\| \beta^{(0)}\|_1 + \| \beta^{(1)}\|_1}{M} \right)\\
 &+ \sum_{a \in \{0, 1\}} \left| \frac{1}{M}\left(\sum_{n \in \Mc^{(a)}} \Ex[Y_{n, t^*}\mid \bA, \bU] 
+ \sum_{n \in \Ic^{(a)}} \beta^{(a)}_n \Ex\left[ Y_{n, t^*} \mid \bA, \bU \right]\right)
- \frac{1}{M} \left(\sum_{n \in \Ic^{(a)}} Y_{n, t^*} 
+ \sum_{n \in \Ic^{(a)}} \hbeta^{(a)}_j Y_{n, t^*} \right) \right|.\label{eq:ate_est_target}
\end{align}
From \eqref{eq:ate_est_target}, it suffices to bound the following terms for $a \in \{0, 1\}$, 
\begin{align}
\left| \frac{1}{M} \left(\sum_{n \in \Mc^{(a)}} Y_{n, t^*} \right)  
- \frac{1}{M}\left(\sum_{n \in \Mc^{(a)}} \Ex[Y_{n, t^*} \mid \bA, \bU ] \right) \right|, \label{eq:ate_est_target_2.1}
\\ \left| \frac{1}{M} \left(\sum_{n \in \Ic^{(a)}} \hbeta^{(a)}_n Y_{n, t^*} \right)  
- \frac{1}{M}\left(\sum_{n \in \Ic^{(a)}} \beta^{(a)}_n \Ex\left[ Y_{n, t^*} \mid \bA, \bU \right]\right) \right|.\label{eq:ate_est_target_3}
\end{align}
For \eqref{eq:ate_est_target_2.1}, using Assumptions \ref{assumption:DGP},
\begin{align}\label{eq:ate_est_target_4}
\frac{1}{M} \left(\sum_{n \in \Mc^{(a)}} Y_{n, t^*} \right) 
- \frac{1}{M}\left(\sum_{n \in \Mc^{(a)}} \Ex[Y_{n, t^*}  \mid \bA, \bU] \right) 
&= \frac{1}{M} \left(\sum_{n \in \Mc^{(a)}} \varepsilon^{(a)}_{n, t^*} \right)
\end{align}
For \eqref{eq:ate_est_target_3}, Using the definition of $\Delta_{\beta^{(a)}}$ we have
\begin{align}
&\frac{1}{M} \left(\sum_{n \in \Ic^{(a)}} \hbeta^{(a)}_n Y_{n, t^*} \right)  
- \frac{1}{M}\left(\sum_{n \in \Ic^{(a)}} \beta^{(a)}_n \Ex\left[ Y_{n, t^*}  \mid \bA, \bU \right]\right) 
\\&= \frac{1}{M}\left(
\ldot{\beta^{(a)}}{\ \varepsilon_{\Ic^{(a)}} }  
+ \ldot{\Delta_{\beta^{(a)}}}{\ \varepsilon_{\Ic^{(a)}} } 
+ \ldot{\Delta_{\beta^{(a)}}}{\ \Ex\left[ Y_{\Ic^{(a)}} \mid \bA, \bU \right]} \right) \label{eq:ate_est_target_5}
\end{align}
From \eqref{eq:ate_est_target}, \eqref{eq:ate_est_target_4}, \eqref{eq:ate_est_target_5} we can write
\begin{align}
|&\widehat{\mathsf{ATE}}_{\Mc}-\Ex\left[\mathsf{ATE}_{\Mc} \mid \bU \right] | \le \textsf{Bias} + \textsf{Variance}
\end{align}
where
\begin{align}
\textsf{Bias} &= \Delta_E \left(1 + \frac{\| \beta^{(0)}\|_1 + \| \beta^{(1)}\|_1}{M} \right) + \sum_{a \in \{0, 1\}} \frac{1}{M} \left( \ldot{\Delta_{\beta^{(a)}}}{\ \varepsilon_{\Ic^{(a)}} } 
+ \ldot{\Delta_{\beta^{(a)}}}{\ \Ex\left[ Y_{\Ic^{(a)}} \mid \bA, \bU \right]} \right)
\\ \textsf{Variance} &= \sum_{a \in \{0, 1\}} \frac{1}{M} \left(\sum_{n \in \Mc^{(a)}} \varepsilon^{(a)}_{n, t^*} \right) +\frac{1}{M}\ldot{\beta^{(a)}}{\ \varepsilon_{\Ic^{(a)}} }
\end{align}
We now further bound the $\textsf{Bias}$ and $\textsf{Variance}$ terms.

\noindent{\em Bounding $\textsf{Variance}$.}

We consider the two terms in $\textsf{Variance}$ separately. 
To bound these two terms, we apply Hoeffding's inequality, which we restate next.
\begin{lemma}[Hoeffding's inequality, e.g., \citeauthor{vershynin_2018}, \citeyear{vershynin_2018}]\label{lemma:hoeffdings}
Let $X_1, \dots, X_N$ be independent mean zero sub-Gaussian random variables, and $a = (a_1, \dots, a_N) \in \Rb^N$.
Then, for every $t \ge 0,$ we have
\begin{align}
\Pb\left( \left| \sum^N_{n = 1} a_n X_n \right| \ge t \right) \le 2 \exp\left( - \frac{Ct^2}{K^2 \|a \|_2^2}\right)
\end{align}
where $K = \max \{\| X_n \|_{\psi_2}\}_{n=1}^N$.
\end{lemma}
Using Assumptions \ref{assumption:DGP}, and \ref{assumption:sub-gaussian_noise}, and applying Hoeffding's inequality from Lemma \ref{lemma:hoeffdings} (with $X_n =  \varepsilon^{(a)}_{n, t^*}$, $a_n = 1$, $K = \sigma_{\max}, t = \sigma_{\max}\sqrt{M_a}$), we have that \eqref{eq:ate_est_target_4} is bounded by
\begin{align}
\frac{1}{M} \left(\sum_{n \in \Mc^{(a)}} \varepsilon^{(a)}_{n, t} \right) 
= O_p\left(  \frac{ \sigma_\max \sqrt{M_a} }{M}\right) \label{eq:ate_est_final_1}
\end{align}
Similarly, we have 
\begin{align}
\frac{1}{M}\ldot{\beta^{(a)}}{\ \varepsilon_{\Ic^{(a)}} }
= O_p\left(\frac{\sigma_\max\|\beta^{(a)} \|_2}{M}\right) \label{eq:ate_est_final_2}
\end{align}

\noindent{\em Bounding $\textsf{Bias}$.}

Again, using Assumptions \ref{assumption:DGP} and \ref{assumption:sub-gaussian_noise}, and using Lemma \ref{lemma:hoeffdings}, we have
\begin{align}
&\frac{1}{M}\ldot{\Delta_{\beta^{(a)}}}{\ \varepsilon_{\Ic^{(a)}}}
= O_p\left( \frac{\sigma_\max\|\Delta_{\beta^{(a)}}\|_2 }{M}\right) \label{eq:ate_est_final_3}
\end{align}
Collecting terms completes the proof.


\subsection{Proof of Theorem \ref{thm:consistency_summary}.}

\vspace{2mm}
\subsubsection{Bounding linear parameter estimation error of PCR.}

We first state and prove two key propositions required to establish Theorem \ref{thm:consistency_summary} that bound $\Delta_{\beta^{(a)}}$.

\begin{proposition}\label{prop:pcr_est}
Let the conditions of Theorem \ref{thm:ate_idenfication}, and Assumptions  \ref{assumption:sub-gaussian_noise}, \ref{assumption:pcr_well_balanced} hold.
Suppose we estimate $\hbeta^{(a)}$ via PCR (i.e., \eqref{eq:PCR_step1} and  \eqref{eq:PCR_step2}) and $k = \bar{r}$.
Then with probability $1 - O((N_a \bar{T})^{-10})$
\begin{align}
\|\Delta_{\beta^{(a)}}\|_2 \le 
C \cdot \sigma_\max^3 \cdot \ln^3(\bar{T} N_a) \cdot \left[\left\| \beta^{(a)} \right\|_2 \cdot \left( \frac{r}{\min(\sqrt{\bar{T}}, \sqrt{N_a})} + r \Delta_E \right) + \frac{M_{1-a} \sqrt{r} \Delta_E}{\sqrt{N_a}} \right].
\end{align}
\end{proposition}

\noindent
{\bf Proof of Proposition \ref{prop:pcr_est}.}

\begin{table}[h]
\centering
\begin{tabular}{||c c||} 
\hline
Notation of \cite{agarwal2021causal} & Our Notation \\ [1ex] 
\hline
$\bY$ & $\frac{\bY}{M_{1-a}}$  \\ [1ex]
\hline
$\bX$ & $\bX$ \\  [1ex]
\hline
$\bZ$ &  $\bZ$  \\ [1ex]
\hline
$\bX^{(lr)}$ &  $\bX^{lr}$  \\ [1ex]
\hline 
$n$ & $\bar{T}$ \\
\hline 
$p$ & $N_a$ \\
\hline 
$\bbeta^*$ & $\frac{\beta^{(a)}}{M_{1-a}}$ \\
\hline
$\hat{\bbeta}$ & $\frac{\hbeta^{(a)}}{M_{1-a}}$ \\
\hline
$\Delta_E$ & $\Delta_E$ \\
\hline 
$r$ & $\bar{r} \ (\le r)$ \\
\hline 
$\phi^{(lr)}$ & $\frac{\phi^{lr}}{M_{1-a}}$ \\ [1ex]
\hline
$\varepsilon$ & $\frac{\bar{\varepsilon}}{M_{1-a}}$ \\ [1ex]
\hline
$\Bar{A}$ & $C$ \\ 
\hline
$\Bar{K}$ & 0  \\ [1ex] 
\hline
$K_{a}, \kappa, \Bar{\sigma}$ & $C\sigma_{\max}$ \\
\hline 
$\rho_{min}$ & 1 \\
\hline 
\end{tabular}
\caption{A summary of the main notational differences between our setting and that of \cite{agarwal2021causal}.}
\label{tab:notational_changes}
\end{table}

Using \eqref{eq:PCR_remark_0}, \eqref{eq:PCR_remark_1}, \eqref{eq:PCR_remark_2}, and \eqref{eq:PCR_remark_3}, we have reduced our problem of estimating $\beta^{\Ic_t^{(a)}}$ to that of linear regression where:
(i) the covariates are noisily observed (i.e., error-in-variables regression), i.e. \eqref{eq:PCR_remark_3} holds;
(ii) the noiseless covariate matrix has an approximate low-rank representation, i.e. \eqref{eq:PCR_remark_2} holds;
(ii) an approximate linear model holds between the approximate low-rank representation of the noiseless covariates and the response variable, i.e. \eqref{eq:PCR_remark_1} holds.
We observe that bounding $\|\Delta_{\beta^{(a)}}\|_2 $ in such a setting is exactly the setup considered in Proposition E.3 of \cite{agarwal2021causal}, where they also analyze PCR.
We match notation with that of \cite{agarwal2021causal} as seen in Table \ref{tab:notational_changes}.
We then get
\begin{align}
\left\|\frac{\Delta_{\beta^{(a)}}}{M_{1 - a}}\right\|_2 
&\le  C \cdot (\sigma_\max)(2\sigma_\max) \cdot \sigma_\max \cdot \ln^3(\bar{T} N_a) \cdot \sqrt{r} \cdot \left( \frac{\|\phi^{\text{lr}}\|_2}{\sqrt{N_a \bar{T}}} + \sqrt{r} \cdot \left\| \frac{\beta^{(a)}}{M_{1 - a}} \right\|_2 \cdot \left( \frac{1}{\sqrt{\bar{T}}} + \frac{1}{\sqrt{N_a}} +  \Delta_E \right)\right) \label{eq:PCR_proof_4}
\end{align}
Using $\|\phi^{\text{lr}}\|_2  \le \Delta_E \sqrt{\bar{T}}$ and simplifying \eqref{eq:PCR_proof_4} completes the proof

\begin{proposition}\label{prop:pcr_est_proj}
Let the conditions of Proposition \ref{prop:pcr_est} hold.
Let $\mathsf{Proj}$ be the projection operator onto the rowspace of $\bX^{\text{lr}}$, i.e., $\mathsf{Proj} = \bV_r \bV_r^T$, where $\bV_r$ are the right singular vectors of $\bX^{\text{lr}}$.
Then with probability $1 - O((N_a \bar{T})^{-10})$
\begin{align}
&\|\mathsf{Proj}(\Delta_{\beta^{(a)}})\|_2 \ \le
\\&  
\quad C \cdot \sigma_\max^4 \cdot \ln^{9/2}(\bar{T} N_a) \cdot 
\left\{ \left\| \beta^{(a)} \right\|_2 \cdot \left( \frac{r^{3/2}}{\min(\bar{T}, N_a)} + \frac{r^{3/2} \Delta_E}{\min(\sqrt{\bar{T}}, \sqrt{N_a})} + r^{3/2} \Delta^2_E \right) \right\},
\\& + C \cdot \sigma_\max^4 \cdot \ln^{9/2}(\bar{T} N_a) \cdot  \left\{ \left\| \beta^{(a)} \right\|_1 \cdot \left( \frac{\sqrt{r}}{\left\| \frac{\beta^{(a)}}{M_{1 - a}} \right\|^{1/2}_1 \bar{T}^{\frac{1}{4}} \sqrt{N_a}} +
\frac{r}{\min(\bar{T}, N_a)} 
+ \frac{r \Delta_E}{\sqrt{N_a}} \right)
\right\},
\\& + C \cdot \sigma_\max^4 \cdot \ln^{9/2}(\bar{T} N_a) \cdot M_{1-a} \cdot \left\{\frac{\sqrt{r} \Delta_E}{\sqrt{N_a}} + \frac{r \Delta^2_E}{\sqrt{N_a}} \right\}.
\end{align}
\end{proposition}

\noindent
{\bf Proof of Proposition \ref{prop:pcr_est_proj}.}
As in the proof of Proposition \ref{prop:pcr_est}, we use \eqref{eq:PCR_remark_1}, \eqref{eq:PCR_remark_2}, \eqref{eq:PCR_remark_3} and observe that bounding $\|\textsf{Proj}(\Delta_{\beta^{(a)}})\|_2$ in such a setting is exactly the setup considered in Corollary E.1 of \cite{agarwal2021causal}, where they also analyze PCR.
Matching notation with that of \cite{agarwal2021causal},
\footnote{
The additional notation compared to Proposition \ref{prop:pcr_est} that needs to be matched here is $\bV_r \bV^T_r = \textsf{Proj}(\cdot)$, where $\bV_r \bV^T_r$ is the notation used in \cite{agarwal2021causal}.
}
we get
\begin{align}
\left\|\frac{\textsf{Proj}(\Delta_{\beta^{(a)}})}{M_{1-a}} \right\|_2 
&\le  C \cdot (\sigma_\max)(2\sigma_\max)^2 \cdot \sigma_\max \cdot \ln^{9/2}(\bar{T} N_a) \cdot \sqrt{r} \cdot \Big[ (A) + (B) + (C)\Big]
\end{align}
where
\begin{align}
(A) &\coloneqq \frac{1}{\sqrt{\bar{T}}} \| \phi^{\text{lr}} \|_2 \left( \frac{1}{\sqrt{N_a}} + \frac{\sqrt{r}}{N_a} + \frac{\sqrt{r}}{\sqrt{\bar{T} N_a}} + \frac{\sqrt{r}}{\sqrt{N_a}} \Delta_E  \right)
\\ (B) &\coloneqq  \left\| \frac{\beta^{(a)}}{M_{1 - a}} \right\|_1 \left( \frac{\bar{T}^{1/4}}{\left\| \frac{\beta^{(a)}}{M_{1 - a}} \right\|^{1/2}_1 \sqrt{\bar{T} N_a}} +
\frac{\sqrt{r}}{\sqrt{\bar{T} N_a}} +
\frac{\sqrt{r}}{N_a}+
\frac{\sqrt{r}}{\sqrt{N_a}} \Delta_E  \right)
\\ (C) &\coloneqq \left\| \frac{\beta^{(a)}}{M_{1 - a}} \right\|_2 \cdot r \cdot \left( 
\frac{1}{\bar{T}} +
\frac{1}{N_a} +
\frac{1}{\sqrt{\bar{T} N_a}} +
\left( \frac{1}{\sqrt{\bar{T}}} + \frac{1}{\sqrt{N_a}} \right) \Delta_E  +
\Delta^2_E  \right)
\end{align}

Using $\|\phi^{\text{lr}}\|_2 \le \sqrt{\bar{T}}\Delta_E$ and $r \le \min(N_a, \bar{T})$, we have 
\begin{align}
(A) &\le \frac{\Delta_E}{\sqrt{N_a}} + \frac{\sqrt{r} \Delta^2_E}{\sqrt{N_a}}.
\end{align}
Simplifying (B) and (C), we have
\begin{align}
(B) &\le \left\| \frac{\beta^{(a)}}{M_{1 - a}} \right\|_1 \left( \frac{1}{\left\| \frac{\beta^{(a)}}{M_{1 - a}} \right\|^{1/2}_1 \bar{T}^{\frac{1}{4}} \sqrt{N_a}} +
\frac{\sqrt{r}}{\min(N_a, \bar{T})} 
+ \frac{\sqrt{r} \Delta_E}{\sqrt{N_a}} \right)
\\ (C) &\le \left\| \frac{\beta^{(a)}}{M_{1 - a}} \right\|_2 \cdot r \cdot \left( \frac{1}{\min(\bar{T}, N_a)} + \frac{\Delta_E}{\min(\sqrt{\bar{T}}, \sqrt{N_a})} + \Delta^2_E \right)
\end{align}
Collecting the various bounds completes this section.

\subsubsection{General conditions for ATE consistency.}
For simplicity, we suppress the conditioning on $\bA, \bU$ in the remainder of the proof.
\begin{proposition}\label{prop:ate_estimation_summary}
Let the conditions of Proposition \ref{prop:pcr_est} and Assumption \ref{assumption:pcr_subspace} hold.
For $a \in \{0, 1\}$, assume $$\|\beta^{(a)}\|_2 = O\left( \frac{M_{1-a}}{N^{w}_a} \right),$$ where $0 \le w \le \frac{1}{2}$.
Then,
\begin{align}
&\widehat{\mathsf{ATE}_{\Mc}} - \Ex[\mathsf{ATE}_{\Mc} \mid \bU]  
\\& = \sum_{a \in \{0, 1\}} C \cdot  \Delta_E \left( \frac{M_{1 - a} \cdot N_a^{0.5 - w}}{M} \right),
\\& \quad + C \cdot \frac{1}{\sqrt{M}},
\\& \quad + \sum_{a \in \{0, 1\}} C \cdot \frac{M_{1-a} \cdot N_a^{- w}}{M},
\\& \quad + \sum_{a \in \{0, 1\}}C \cdot \frac{M_{1-a}}{M N^w_a} \cdot  \sigma_\max^3 \cdot \ln^3(\bar{T} N_a) \cdot \left[r \left( \frac{1}{\min(\sqrt{\bar{T}}, \sqrt{N_a})} +  \Delta_E \right) \right],
\\& \quad + \sum_{a \in \{0, 1\}} C \cdot \frac{M_{1-a}}{M} \cdot N_a^{1/2 - w} \cdot \sigma_\max^3 \cdot \ln^3(\bar{T} N_a) \cdot \left[r \left( \frac{1}{\min(\sqrt{\bar{T}}, \sqrt{N_a})} +  \Delta_E \right) \right]\Delta_E,
\\& \quad + \sum_{a \in \{0, 1\}} C \cdot \frac{1}{M} \cdot \sigma_\max^4 \cdot \ln^{9/2}(\bar{T} N_a) \cdot 
\left\{ \frac{M_{1-a}}{N^{w - 0.5}_a} \cdot \left( \frac{r^{3/2}}{\min(\bar{T}, N_a)} + \frac{r^{3/2} \Delta_E}{\min(\sqrt{\bar{T}}, \sqrt{N_a})} + r^{3/2} \Delta^2_E \right) \right\},
\\& \quad + \sum_{a \in \{0, 1\}} C \cdot \frac{\sqrt{N_a} M_{1-a}}{M} \cdot \sigma_\max^4 \cdot \ln^{9/2}(\bar{T} N_a) \cdot  \left\{ \frac{\sqrt{r}}{ \bar{T}^{\frac{1}{4}} N_a^{0.5w + 0.25}} + \frac{ r}{N_a^{w - 0.5} \cdot \min(\bar{T}, N_a)} + \frac{ r \cdot \Delta_E}{N_a^{w}} \right\},
\\& \quad +\sum_{a \in \{0, 1\}} C \cdot \frac{ M_{1-a}}{M} \cdot \sigma_\max^4 \cdot \ln^{9/2}(\bar{T} N_a) \cdot  \left\{\sqrt{r}\Delta_E + r \Delta^2_E \right\}.
\end{align}
\end{proposition}

\noindent
{\bf Proof of Proposition \ref{prop:ate_estimation_summary}.}

From Proposition \ref{prop:ate_estimation}, we have that
\begin{align}
&\widehat{\mathsf{ATE}_{\Mc}} - \Ex[\mathsf{ATE}_{\Mc} \mid \bU] 
\\&\le  C  \Delta_E \left(1 + \frac{\| \beta^{(0)}\|_1 + \| \beta^{(1)}\|_1}{M} \right) 
+ O_p\left( \sum_{a \in \{0, 1\}} \frac{\sigma_\max \left(\sqrt{M_a} +  \|\beta^{(a)} \|_2 + \|\Delta_{\beta^{(a)}}\|_2\right) +  \ldot{\Delta_{\beta^{(a)}}  }{\Ex\left[ Y_{\Ic^{(a)}} \right]}}{M} \right) 
\end{align}
We consider the various terms on the right-hand side above separately.

\vspace{2mm}
\noindent{\em 1. Bounding the $\Delta_E \left(1 + \frac{\| \beta^{(0)}\|_1 + \| \beta^{(1)}\|_1}{M} \right)$ term.}

\noindent
Given the assumption that $ \| \beta^{(a)} \|_2 = O\left( \frac{M_{1-a}}{N^w_a} \right)$, we have $\| \beta^{(a)} \|_1 = O\left( \frac{M_{1-a} N_a^{0.5}}{N_a^w} \right).$
Therefore
\begin{align}
\Delta_E \left(\frac{\| \beta^{(0)}\|_1 + \| \beta^{(1)}\|_1}{M} \right) 
&= C \cdot  \Delta_E \left( \frac{M_0 \cdot N_1^{0.5 - w} + M_1 \cdot N_0^{0.5 - w}}{M} \right) 
\\&= \sum_{a \in \{0, 1\}} C \cdot  \Delta_E \left( \frac{M_{1 - a} \cdot N_a^{0.5 - w}}{M} \right)
\label{eq:summary_4}
\end{align}

\noindent{\em 2. Bounding the $\sum_{a \in \{0, 1\}}  \frac{\sigma_\max (\sqrt{M_a} +  \|\beta^{(a)} \|_2)}{M}$ term.}

\noindent
Note that since $M_a < M$,
\begin{align}
\sum_{a \in \{0, 1\}}  \frac{\sigma_\max \sqrt{M_a}}{M}
= O\left(\frac{1}{\sqrt{M}}\right). \label{eq:summary_9}
\end{align}
Further,
\begin{align}
\sum_{a \in \{0, 1\}}  \frac{\|\beta^{(a)} \|_2}{M} 
= C \cdot \frac{M_0 \cdot N_1^{- w} + M_1 \cdot N_0^{- w}}{M} 
= \sum_{a \in \{0, 1\}} C \cdot \frac{M_{1-a} \cdot N_a^{- w}}{M} 
\label{eq:summary_10}
\end{align}

\noindent{\em 3. Bounding the $\sum_{a \in \{0, 1\}}  \frac{\sigma_\max \left\|\Delta_{\beta^{(a)}}\right\|_2}{M}$ term.}

\noindent
Using Proposition \ref{prop:pcr_est} and $ \| \beta^{(a)} \|_2 = O\left( \frac{M_{1-a}}{N^w_a} \right)$, we have
\begin{align}
&\sum_{a \in \{0, 1\}}\frac{\|\Delta_{\beta^{(a)}}\|_2}{M}
\\ &\le \sum_{a \in \{0, 1\}}\frac{1}{M} \cdot C \cdot \sigma_\max^3 \cdot \ln^3(\bar{T} N_a) \cdot \left[\left\| \beta^{(a)} \right\|_2 \cdot \left( \frac{r}{\min(\sqrt{\bar{T}}, \sqrt{N_a})} + r \Delta_E \right) + \frac{M_{1-a} \sqrt{r} \Delta_E}{\sqrt{N_a}} \right],
\\ &\le \sum_{a \in \{0, 1\}}\frac{M_{1-a}}{M} \cdot C \cdot \sigma_\max^3 \cdot \ln^3(\bar{T} N_a) \cdot \left[\frac{1}{N^w_a} \cdot \left( \frac{r}{\min(\sqrt{\bar{T}}, \sqrt{N_a})} + r \Delta_E \right) + \frac{\sqrt{r} \Delta_E}{\sqrt{N_a}} \right],
\\ &\le \sum_{a \in \{0, 1\}} C \cdot \frac{M_{1-a}}{M N^w_a} \cdot  \sigma_\max^3 \cdot \ln^3(\bar{T} N_a) \cdot \left[r \left( \frac{1}{\min(\sqrt{\bar{T}}, \sqrt{N_a})} +  \Delta_E \right) \right],
\label{eq:summary_2}
\end{align}
where in the third inequality we have used that $w \le \frac{1}{2}$.

\noindent{\em 4. Bounding the $\sum_{a \in \{0, 1\}}  \frac{\ldot{\Delta_{\beta^{(a)}} }{\Ex\left[ Y_{\Ic^{(a)}} \right]}}{M}$ term.}

\noindent
From Definition \ref{def:linear_factor_approx}, we have that $\Ex[Y^{(a)}_{j, t^*}] = \ldot{\lambda_j}{\rho_{t^*, a}} + \eta^{(a)}_{j, t^*}$.
Hence
\begin{align}
\sum_{a \in \{0, 1\}} \left| \ldot{\Delta_{\beta^{(a)}} }{\Ex\left[ Y_{\Ic^{(a)}} \right]} \right|
&= \sum_{a \in \{0, 1\}} \left| \ldot{\Delta_{\beta^{(a)}} }{[\ldot{\lambda_j}{\rho_{t^*, a}} + \eta^{(a)}_{j, t^*}]_{n \in \Ic^{(a)}}} \right|
\\ &= \sum_{a \in \{0, 1\}}  \left| \ldot{\Delta_{\beta^{(a)}} }{[\ldot{\lambda_j}{\rho_{t^*, a}}]_{n \in \Ic^{(a)}}} + \ldot{\Delta_{\beta^{(a)}} }{[\eta^{(a)}_{j, t^*}]_{n \in \Ic^{(a)}}} \right|
\\ &\le \sum_{a \in \{0, 1\}} \left| \ldot{\Delta_{\beta^{(a)}} }{[\ldot{\lambda_j}{\rho_{t^*, a}}]_{n \in \Ic^{(a)}}} \right| + \|\Delta_{\beta^{(a)}}\|_2 \| [\eta^{(a)}_{j, t^*}]_{n \in \Ic^{(a)}} \|_2 
\\ &\le \sum_{a \in \{0, 1\}} \left| \ldot{\Delta_{\beta^{(a)}} }{[\ldot{\lambda_j}{\rho_{t^*, a}}]_{n \in \Ic^{(a)}}} \right| + \|\Delta_{\beta^{(a)}}\|_2 \sqrt{N_a} \Delta_E
\end{align}
Using Assumption \ref{assumption:pcr_subspace}, we have
\begin{align}
\left| \ldot{\Delta_{\beta^{(a)}} }{[\ldot{\lambda_j}{\rho_{t^*, a}}]_{n \in \Ic^{(a)}}} \right|
&= \left| \ldot{\Delta_{\beta^{(a)}} }{\mathsf{Proj}([\ldot{\lambda_j}{\rho_{t^*, a}}]_{n \in \Ic^{(a)}}}) \right|
\\&= \left| \ldot{\mathsf{Proj}(\Delta_{\beta^{(a)}})}{[\ldot{\lambda_j}{\rho_{t^*, a}}]_{n \in \Ic^{(a)}}} \right|
\\& \le \| \mathsf{Proj}(\Delta_{\beta^{(a)}}) \|_2 \| [\ldot{\lambda_j}{\rho_{t^*, a}}]_{n \in \Ic^{(a)}} \|_2
\\& \le C \| \mathsf{Proj}(\Delta_{\beta^{(a)}}) \|_2 \sqrt{N_a},
\end{align}
where recall $\mathsf{Proj} = \bV_r \bV_r^T$, and $\bV_r$ are the right singular vectors of $\bX^{\text{lr}}$.

Hence, we have
\begin{align}
\frac{1}{M}\sum_{a \in \{0, 1\}} \left| \ldot{\Delta_{\beta^{(a)}} }{\Ex\left[ Y_{\Ic^{(a)}}  \right]} \right|
&\le \sum_{a \in \{0, 1\}} \frac{1}{M}\|\Delta_{\beta^{(a)}}\|_2 \sqrt{N_a} \Delta_E + \frac{C}{M}  \| \mathsf{Proj}(\Delta_{\beta^{(a)}}) \|_2 \sqrt{N_a}  
\end{align}
We bound each term above separately.

\noindent{\em 4a. Bounding the $ \sum_{a \in \{0, 1\}} \frac{1}{M}\|\Delta_{\beta^{(a)}}\|_2 \sqrt{N_a} \Delta_E$ term.}
For the first term, by applying a similar logic used to derive \eqref{eq:summary_2}, we have that
\begin{align}
&\sum_{a \in \{0, 1\}} \frac{1}{M}\|\Delta_{\beta^{(a)}}\|_2 \sqrt{N_a} \Delta_E 
\\&\le \sum_{a \in \{0, 1\}}\frac{M_{1-a}}{M N^w_a} \cdot C \cdot \sigma_\max^3 \cdot \ln^3(\bar{T} N_a) \cdot \left[r \left( \frac{1}{\min(\sqrt{\bar{T}}, \sqrt{N_a})} +  \Delta_E \right) \right]\sqrt{N_a}\Delta_E
\\&\le \sum_{a \in \{0, 1\}}\frac{M_{1-a}}{M} \cdot N_a^{1/2 - w} \cdot C \cdot \sigma_\max^3 \cdot \ln^3(\bar{T} N_a) \cdot \left[r \left( \frac{1}{\min(\sqrt{\bar{T}}, \sqrt{N_a})} +  \Delta_E \right) \right]\Delta_E
\label{eq:summary_5}
\end{align}

\noindent{\em 4b. Bounding the $\sum_{a \in \{0, 1\}} \frac{C}{M} \| \mathsf{Proj}(\Delta_{\beta^{(a)}}) \|_2 \sqrt{N_a}$ term.}

\noindent
Using Proposition \ref{prop:pcr_est_proj} and $ \| \beta^{(a)} \|_2 = O\left( \frac{M_{1-a}}{N^w_a} \right)$, we have
\begin{align}
&\sum_{a \in \{0, 1\}} \frac{C}{M} \| \mathsf{Proj}(\Delta_{\beta^{(a)}}) \|_2 \sqrt{N_a}
\\& \le \sum_{a \in \{0, 1\}}
\quad \frac{C\sqrt{N_a}}{M} \cdot \sigma_\max^4 \cdot \ln^{9/2}(\bar{T} N_a) \cdot 
\left\{ \left\| \beta^{(a)} \right\|_2 \cdot \left( \frac{r^{3/2}}{\min(\bar{T}, N_a)} + \frac{r^{3/2} \Delta_E}{\min(\sqrt{\bar{T}}, \sqrt{N_a})} + r^{3/2} \Delta^2_E \right) \right\},
\\& \quad \quad + \frac{C\sqrt{N_a}}{M} \cdot \sigma_\max^4 \cdot \ln^{9/2}(\bar{T} N_a) \cdot  \left\{ \left\| \beta^{(a)} \right\|_1 \cdot \left( \frac{\sqrt{r}}{\left\| \frac{\beta^{(a)}}{M_{1 - a}} \right\|^{1/2}_1 \bar{T}^{\frac{1}{4}} \sqrt{N_a}} +
\frac{r}{\min(\bar{T}, N_a)} + \frac{r \Delta_E}{\sqrt{N_a}} \right) \right\},
\\& \quad \quad + \frac{C\sqrt{N_a}}{M} \cdot \sigma_\max^4 \cdot \ln^{9/2}(\bar{T} N_a) \cdot M_{1-a} \cdot  \left\{\frac{\sqrt{r}\Delta_E}{\sqrt{N_a}} + \frac{r \Delta^2_E}{\sqrt{N_a}} \right\}. \label{eq:summary_3}
\end{align}

We bound the three terms on the r.h.s above separately. 

\begin{enumerate}
\item First term of \eqref{eq:summary_3}.
\begin{align}
&\sum_{a \in \{0, 1\}}
\frac{C\sqrt{N_a}}{M} \cdot \sigma_\max^4 \cdot \ln^{9/2}(\bar{T} N_a) \cdot 
\left\{ \left\| \beta^{(a)} \right\|_2 \cdot \left( \frac{r^{3/2}}{\min(\bar{T}, N_a)} + \frac{r^{3/2} \Delta_E}{\min(\sqrt{\bar{T}}, \sqrt{N_a})} + r^{3/2} \Delta^2_E \right) \right\}
\\& \le \sum_{a \in \{0, 1\}}
\quad \frac{C}{M} \cdot \sigma_\max^4 \cdot \ln^{9/2}(\bar{T} N_a) \cdot 
\left\{ \frac{M_{1-a}}{N^{w - 0.5}_a} \cdot \left( \frac{r^{3/2}}{\min(\bar{T}, N_a)} + \frac{r^{3/2} \Delta_E}{\min(\sqrt{\bar{T}}, \sqrt{N_a})} + r^{3/2} \Delta^2_E \right) \right\}. \label{eq:summary_6}
\end{align}

\item Second term of \eqref{eq:summary_3}.
\begin{align}
&\sum_{a \in \{0, 1\}} \frac{C\sqrt{N_a}}{M} \cdot \sigma_\max^4 \cdot \ln^{9/2}(\bar{T} N_a) \cdot  \left\{ \left\| \beta^{(a)} \right\|_1 \cdot \left( \frac{\sqrt{r}}{\left\| \frac{\beta^{(a)}}{M_{1 - a}} \right\|^{1/2}_1 \bar{T}^{\frac{1}{4}} \sqrt{N_a}} +
\frac{r}{\min(\bar{T}, N_a)} + \frac{r \Delta_E}{\sqrt{N_a}} \right) \right\}
\\ &\le \sum_{a \in \{0, 1\}} \frac{C\sqrt{N_a}}{M} \cdot \sigma_\max^4 \cdot \ln^{9/2}(\bar{T} N_a) \cdot  \left\{ \frac{\sqrt{r} \left\| \beta^{(a)} \right\|^{0.5}_1 M_{1 - a}^{0.5}}{ \bar{T}^{\frac{1}{4}} \sqrt{N_a}} + \frac{ \left\| \beta^{(a)} \right\|_1 r}{\min(\bar{T}, N_a)} + \frac{ \left\| \beta^{(a)} \right\|_1 r \Delta_E}{\sqrt{N_a}} \right\},
\\ &\le \sum_{a \in \{0, 1\}} \frac{C\sqrt{N_a}}{M} \cdot \sigma_\max^4 \cdot \ln^{9/2}(\bar{T} N_a) \cdot  \left\{ \frac{\sqrt{r} \cdot M_{1-a} }{ \bar{T}^{\frac{1}{4}} N_a^{0.5w + 0.25}} + \frac{ M_{1-a} \cdot r}{N_a^{w - 0.5} \cdot \min(\bar{T}, N_a)} + \frac{ M_{1-a} \cdot r \cdot \Delta_E}{N_a^{w}} \right\},
\\ &\le \sum_{a \in \{0, 1\}} \frac{C\sqrt{N_a} M_{1-a}}{M} \cdot \sigma_\max^4 \cdot \ln^{9/2}(\bar{T} N_a) \cdot  \left\{ \frac{\sqrt{r} }{ \bar{T}^{\frac{1}{4}} N_a^{0.5w + 0.25}} + \frac{ r}{N_a^{w - 0.5} \cdot \min(\bar{T}, N_a)} + \frac{ r \cdot \Delta_E}{N_a^{w}} \right\}. \label{eq:summary_7}
\end{align}

\item Third term of \eqref{eq:summary_3}.
\begin{align}
&\sum_{a \in \{0, 1\}} \frac{C\sqrt{N_a}}{M} \cdot \sigma_\max^4 \cdot \ln^{9/2}(\bar{T} N_a) \cdot M_{1-a} \cdot  \left\{\frac{\sqrt{r}\Delta_E}{\sqrt{N_a}} + \frac{r \Delta^2_E}{\sqrt{N_a}} \right\}
\\ &= \sum_{a \in \{0, 1\}} C \cdot \frac{ M_{1-a}}{M} \cdot \sigma_\max^4 \cdot \ln^{9/2}(\bar{T} N_a) \cdot  \left\{\sqrt{r}\Delta_E + r \Delta^2_E \right\}
\label{eq:summary_8}
\end{align}
\end{enumerate}

\noindent{\em Summarizing all terms.}

Using \eqref{eq:summary_4}, \eqref{eq:summary_9}, \eqref{eq:summary_10}, \eqref{eq:summary_5}, \eqref{eq:summary_3}, \eqref{eq:summary_6}, \eqref{eq:summary_7}, \eqref{eq:summary_8}, we complete the proof of the proposition.

\subsubsection{Finishing proof of Theorem \ref{thm:consistency_summary}.}

Recall from Proposition \ref{prop:factor_model_approx}, we have that
\begin{align}\label{eq:summary_0}
r \le C \cdot \delta^{-q}, \quad \Delta_E \le C \cdot \delta^S.
\end{align}

{\bf $\textsf{ATT}$ consistency.}

For estimation of $\textsf{ATT}$, we have that $M = M_1 = N_1$ and $M_0 = 0$.
Hence, by simplifying the result in Proposition \ref{prop:ate_estimation_summary}, we get that
\begin{align}
&\widehat{\mathsf{ATE}_{\Mc}} - \Ex[\mathsf{ATE}_{\Mc} \mid \bU]  
\\&\le C \cdot  \Delta_E \left(  N_0^{0.5 - w}\right), \label{eq:ATT_summary_1}
\\& \quad + C \cdot \frac{1}{\sqrt{M}}, \label{eq:ATT_summary_2}
\\& \quad + C \cdot  N_0^{- w}, \label{eq:ATT_summary_3}
\\& \quad + C \cdot \frac{1}{N^w_0} \cdot \ln^3(\bar{T} N_0) \cdot \left[r \left( \frac{1}{\min(\sqrt{\bar{T}}, \sqrt{N_0})} +  \Delta_E \right) \right], \label{eq:ATT_summary_4}
\\& \quad + C \cdot N_0^{1/2 - w} \cdot \ln^3(\bar{T} N_0) \cdot \left[r \left( \frac{1}{\min(\sqrt{\bar{T}}, \sqrt{N_0})} +  \Delta_E \right) \right]\Delta_E, \label{eq:ATT_summary_5}
\\& \quad + C \cdot \ln^{9/2}(\bar{T} N_0) \cdot 
\left\{ \frac{1}{N^{w - 0.5}_0} \cdot \left( \frac{r^{3/2}}{\min(\bar{T}, N_0)} + \frac{r^{3/2} \Delta_E}{\min(\sqrt{\bar{T}}, \sqrt{N_0})} + r^{3/2} \Delta^2_E \right) \right\}, \label{eq:ATT_summary_6}
\\& \quad + C \cdot \sqrt{N_0}   \cdot \ln^{9/2}(\bar{T} N_0) \cdot  \left\{ \frac{\sqrt{r}}{ \bar{T}^{\frac{1}{4}} N_0^{0.5w + 0.25}} + \frac{ r}{N_0^{w - 0.5} \cdot \min(\bar{T}, N_0)} + \frac{ r \cdot \Delta_E}{N_0^{w}} \right\},  \label{eq:ATT_summary_7}
\\& \quad + C  \cdot \ln^{9/2}(\bar{T} N_0) \cdot  \left\{\sqrt{r}\Delta_E + r \Delta^2_E \right\}. \label{eq:ATT_summary_8}
\end{align} 

We deal with the seven terms above separately. 

Let $G = \min(N_0, \bar{T})$.
For $\gamma > 0$, take $\delta = \left(\frac{1}{G}\right)^{\gamma / q}$.
Then \eqref{eq:summary_0} implies
\begin{align}\label{eq:summary_1}
r \le C G^\gamma, \quad \Delta_E &\le C \left(\frac{1}{G}\right)^{ \gamma \alpha}.
\end{align}
We set $\gamma = \frac{1}{2\alpha}$ and so we have $r \le C G^{\frac{1}{2\alpha}}$, and that $\Delta_E \le C G^{-0.5}$.

{\em Term \eqref{eq:ATT_summary_1}}.
\begin{align}
&C \cdot  \Delta_E \left(  N_0^{0.5 - w}\right)
\le G^{-\gamma \alpha} N_0^{0.5 - w} 
= G^{-0.5} N_0^{0.5 - w}
= o_p(1)
\end{align}
where in the last line we have used $\alpha > 1$ and the assumption $\frac{N_0^{1 - w}}{\bar{T}^{1 - \frac{1}{2 \alpha}}} = o(1) \implies \frac{N_0^{0.5 - w}}{\bar{T}^{0.5}} = o(1)$; we also use the assumption that $w > 0$. 

{\em Term \eqref{eq:ATT_summary_2} and \eqref{eq:ATT_summary_3}}.

Given the assumption that $M (= N_1), N_0 \to \infty$, and that $w > 0$, 
\begin{align}
C \cdot \frac{1}{\sqrt{M}} &= o_p(1),
\\ C \cdot  N_0^{- w} &= o_p(1).
\end{align}

{\em Term \eqref{eq:ATT_summary_4}}.
\begin{align}
&C \cdot \frac{1}{N^w_0} \cdot \ln^3(\bar{T} N_0) \cdot \left[r \left( \frac{1}{\min(\sqrt{\bar{T}}, \sqrt{N_0})} +  \Delta_E \right) \right]
\\&\le C \cdot \frac{1}{N^w_0} \cdot \ln^3(\bar{T} N_0) \cdot \left[G^{\gamma} \left( \frac{1}{\min(\sqrt{\bar{T}}, \sqrt{N_0})} +  G^{-\gamma \alpha} \right) \right]
\\&\le C \cdot \ln^3(\bar{T} N_0) \cdot \left[G^{\gamma - w - 0.5} + G^{\gamma (1 - \alpha) - w}   \right]
\\&= C \cdot \ln^3(\bar{T} N_0) \cdot \left[G^{0.5(\frac{1}{\alpha} - 1) - w} \right] \label{eq:ATT_normality_helper_1}
\\&= o_p(1)
\end{align}
where in the last line we use the inequality that $w > \frac{1}{2 \alpha} - \frac{1}{2}$.

{\em Term \eqref{eq:ATT_summary_5}}.
\begin{align}
&C \cdot N_0^{1/2 - w} \cdot \ln^3(\bar{T} N_0) \cdot \left[r \left( \frac{1}{\min(\sqrt{\bar{T}}, \sqrt{N_0})} +  \Delta_E \right) \right]\Delta_E
\\&\le C \cdot \ln^3(\bar{T} N_0) \cdot \left[G^{\frac{1}{2 \alpha} - 0.5}   \right] \cdot  G^{-0.5} \cdot N_0^{0.5 - w}  
\\&= o_p(1)
\end{align}
where in the last line we have used $\alpha > 1$ and the assumption $\frac{N_0^{1 - w}}{\bar{T}^{1 - \frac{1}{2 \alpha}}} = o(1) \implies \frac{N_0^{0.5 - w}}{\bar{T}^{0.5}} = o(1)$; we also use the assumption that $w > 0$. 

{\em Term \eqref{eq:ATT_summary_6}}.
\begin{align}
&C \cdot \ln^{9/2}(\bar{T} N_0) \cdot 
\left\{ \frac{1}{N^{w - 0.5}_0} \cdot \left( \frac{r^{3/2}}{\min(\bar{T}, N_0)} + \frac{r^{3/2} \Delta_E}{\min(\sqrt{\bar{T}}, \sqrt{N_0})} + r^{3/2} \Delta^2_E \right) \right\}
\\&\le  C \cdot \ln^{9/2}(\bar{T} N_0) \cdot \left\{ \frac{1}{N^{w - 0.5}_0} \cdot \left( G^{1.5\gamma - 1} + G^{\gamma(1.5 - \alpha) - 0.5} + G^{\gamma(1.5 - 2\alpha)}\right) \right\}
%
%
\\&\le  C \cdot \ln^{9/2}(\bar{T} N_0) \cdot \left\{ \frac{ N_0 ^{0.5 - w} }{G^{1 - \frac{0.75}{\alpha}}}  \right\}
\\&= o_p(1)
\end{align}
where in the last line we have used the fact that $w > \frac{1}{2 \alpha}$, $\alpha > \frac{1}{2}$ and the assumption that $\frac{N_0^{1 - w}}{\bar{T}^{1 - \frac{1}{2 \alpha}}} = o(1)$.

{\em Term \eqref{eq:ATT_summary_7}}.
\begin{align}
&C \cdot \sqrt{N_0}   \cdot \ln^{9/2}(\bar{T} N_0) \cdot  \left\{ \frac{\sqrt{r}}{ \bar{T}^{\frac{1}{4}} N_0^{0.5w + 0.25}} + \frac{ r}{N_0^{w - 0.5} \cdot \min(\bar{T}, N_0)} + \frac{ r \cdot \Delta_E}{N_0^{w}} \right\}
\\&\le C  \cdot \ln^{9/2}(\bar{T} N_0) \cdot  \left\{ \frac{\sqrt{r}}{ \bar{T}^{\frac{1}{4}} N_0^{0.5w - 0.25}} + \frac{r}{N_0^{w -1} \cdot \min(\bar{T}, N_0)} + \frac{ r \cdot \Delta_E}{N_0^{w - 0.5}} \right\}
\\&\le C  \cdot \ln^{9/2}(\bar{T} N_0) \cdot  \left\{ \frac{G^{\frac{1}{4 \alpha}} N_0^{0.25 - 0.5w}}{ \bar{T}^{\frac{1}{4}} } + \frac{N_0^{1 - w}}{G^{1 - \frac{1}{2 \alpha}}} + \frac{N_0^{0.5 - w}}{G^{0.5 - \frac{1}{2\alpha}}} \right\}
\\&= o_p(1)
\end{align}
where in the last line we have used the fact that $w > \frac{1}{2 \alpha}$ and the assumption that $\frac{N_0^{1 - w}}{\bar{T}^{1 - \frac{1}{2 \alpha}}} = o(1)$.

{\em Term \eqref{eq:ATT_summary_8}}.

Using the assumption that $\alpha > \frac{1}{2}$,  we have that 
\begin{align}
&\sqrt{r}\Delta_E \le C G^{0.5 \gamma} \cdot G^{- \gamma \alpha} = G^{\gamma(0.5 - \alpha)} = o_p(1),
\end{align} 
which also implies that $r\Delta^2_E = o_p(1)$.

{\em Completing the proof of $\mathsf{ATT}$ consistency.}

The above inequalities establish that $\widehat{\mathsf{ATT}}  - \Ex[\mathsf{ATT} \mid \bU] = o_p(1)$.
%

{\bf $\textsf{ATU}$ consistency.}

The proof follows in an analogous manner to that of $\mathsf{ATT}$, where we switch the roles of the treated and untreated. 
That is, $M = M_0 = N_0$ and $M_1 = 0$.

{\bf $\textsf{ATE}$ consistency.}

The proof follows in an analogous manner to that of $\mathsf{ATT}$, where now $M_0, M_1 = \Theta(M)$, and we have $M_0 = N_0$, $M_1 = N_1$, $M = N$.

\subsection{Proof of Proposition \ref{prop:beta_norm}.}
For simplicity and without loss of generality, we let the $N_a$ units in $\Ic^{(a)}$ be the indexed as the first $N_a$ units.

Now, given the assumption in the statement of Proposition \ref{prop:beta_norm}, we have that for $k \in [N_{a}^{1 - \theta}]$ there exists $\beta^{n, k} \in \Rb^{N_{a}^{\theta}}$ such that for all $n \in \Mc^{(1-a)}$
\begin{align}
\lambda_n = \sum^{k N_{a}^\theta}_{i = 1 + (k - 1)N_{a}^\theta} \beta^{n, k}_i \lambda_i
\end{align}
and 
\begin{align}
\| \beta^{n, k} \|_2 = O(\sqrt{r}).
\end{align}
Define $\tilde{\beta}^n \in \Rb^{N_a}$ as follows
\begin{align}
\tilde{\beta}^n = \frac{1}{N_{a}^{1 - \theta}}[\beta^{n, 1}, \dots, \beta^{n, N_{a}^{1 - \theta}}].
\end{align}
Note
\begin{align}
\lambda_n = \sum^{N_a}_{i = 1} \tilde{\beta}^n_i \lambda_i.
\end{align}
Then using $1 -  \frac{3}{2\alpha} > \theta   \Longleftrightarrow  \frac{1 - \theta}{2} - \frac{1}{4\alpha} > \frac{1}{2\alpha}$, we have
\begin{align}
\| \tilde{\beta}^n \|_2 
= O\left( \frac{\sqrt{r}}{N_{a}^{\frac{1 - \theta}{2}}} \right) 
= O\left( \frac{N_{a}^{\frac{1}{4\alpha}}}{N_{a}^{\frac{1 - \theta}{2}}} \right)  
= o\left(\frac{1}{N_{a}^{{\frac{1}{2\alpha}}}}\right).
\end{align}
Then define 
\begin{align}
\tilde{\beta}^{\Ic^{(a)}} &= \sum_{n \in \Mc^{(1-a)}} \tilde{\beta}^n,    
\\\implies \sum^{N_a}_{i = 1} \tilde{\beta}^{\Ic^{(a)}}_i \lambda_i &= \sum_{n \in \Mc^{(1-a)}} \sum^{N_a}_{i = 1} \tilde{\beta}^n_i   \lambda_i = \sum_{n \in \Mc^{(1-a)}} \lambda_n = \lambda_{\Mc^{(1-a)}}.
\end{align}

Hence, 
\begin{align}
\| \tilde{\beta}^{\Ic^{(a)}} \|_2 = o\left(\frac{M_{1-a}}{N_{a}^{{\frac{1}{2\alpha}}}}\right).
\end{align}
Since we define $\beta^{(a)}$ to be linear weight with minimum $\ell_2$-norm in Assumption \ref{assumption:linear_span_inclusion}, it follows that 
$\| \beta^{(a)} \|_2 =  o\left(\frac{M_{1-a}}{N_{a}^{{\frac{1}{2\alpha}}}}\right)$.
This completes the proof.

\end{appendix}

\end{document}